\documentclass[pra, aps, reprint, floatfix, tightenlines, twocolumn, longbibliography, superscriptaddress]{revtex4-1}

\usepackage{epsfig}
\usepackage{amsmath,amssymb,amsfonts}
\usepackage{mathrsfs}
\usepackage{bbm}
\usepackage{slashed}
\usepackage{graphicx}
\usepackage{verbatim}
\usepackage[usenames]{color}
\usepackage{mathtools}
\usepackage[ruled,linesnumbered,commentsnumbered]{algorithm2e}
\usepackage{float}
\usepackage{appendix}
\usepackage{booktabs}
\usepackage{relsize}

\usepackage{enumerate}

\usepackage{url}
\usepackage{hyperref}
\hypersetup{
    colorlinks=true,
    linkcolor=blue,
    citecolor=magenta, 
    filecolor=magenta,      
    urlcolor=cyan,
}

\makeatletter
\def\@bibdataout@aps{%
\immediate\write\@bibdataout{%
@CONTROL{%
apsrev41Control%
\longbibliography@sw{%
    ,author="08",editor="1",pages="1",title="0",year="1"%
    }{%
    ,author="08",editor="1",pages="1",title="",year="1"%
    }%
  }%
}%
\if@filesw \immediate \write \@auxout {\string \citation {apsrev41Control}}\fi 
}
\makeatother

\usepackage{mleftright} 

\DeclarePairedDelimiter\abs{\lvert}{\rvert}

\usepackage[nolist]{acronym}

\begin{document}


\begin{acronym}
    \acro{MWPM}{Minimum Weight Perfect Matching}
\end{acronym}

\newacro{MWPM}{Minimum Weight Perfect Matching}
\newacro{CNN}{convolutional neural network}
\newacro{ReLU}{rectified linear unit}
\newacro{SGD}{stochastic gradient descent}
\newacro{RL}{reinforcement learning}
\newacro{DRL}{deep reinforcement learning}
\newacro{NN}{neural network}
\newacro{ML}{machine learning}


\title{Deep Q-learning decoder for depolarizing noise on the toric code}

\author{David Fitzek}
\email[]{davidfi@chalmers.se}
\affiliation{Wallenberg Centre for Quantum Technology, Department of Microtechnology and Nanoscience, Chalmers University of Technology, SE-41296 Gothenburg, Sweden}
\affiliation{Volvo Group Trucks Technology, 405 08 Gothenburg, Sweden}

\author{Mattias Eliasson}
\affiliation{Department of Physics, University of Gothenburg, SE-41296 Gothenburg, Sweden}

\author{Anton Frisk Kockum}
\affiliation{Wallenberg Centre for Quantum Technology, Department of Microtechnology and Nanoscience, Chalmers University of Technology, SE-41296 Gothenburg, Sweden}

\author{Mats Granath}
\email[]{mats.granath@physics.gu.se}
\affiliation{Department of Physics, University of Gothenburg, SE-41296 Gothenburg, Sweden}
	
	
\begin{abstract}

We present an AI-based decoding agent for quantum error correction of depolarizing noise on the toric code. The agent is trained using deep reinforcement learning (DRL), where an artificial neural network encodes the state-action Q-values of error-correcting $X$, $Y$, and $Z$ Pauli operations, occurring with probabilities $p_x$, $p_y$, and $p_z$, respectively. 
By learning to take advantage of the correlations between bit-flip and phase-flip errors, the decoder outperforms the minimum-weight-perfect-matching (MWPM) algorithm, achieving higher success rate and higher error threshold for depolarizing noise ($p_z = p_x = p_y$), for code distances $d\leq 9$. The decoder trained on depolarizing noise also has close to optimal performance for uncorrelated noise and provides functional but sub-optimal decoding for biased noise ($p_z \neq p_x = p_y$). We argue that the DRL-type decoder provides a promising framework for future practical error correction of topological codes, striking a balance between on-the-fly calculations, in the form of forward evaluation of a deep Q-network, and pre-training and information storage. The complete code, as well as ready-to-use decoders (pre-trained networks), can be found in the repository \href{https://github.com/mats-granath/toric-RL-decoder}{github.com/mats-granath/toric-RL-decoder}. 

\end{abstract}

\maketitle


\section{Introduction}

The basic building block of a quantum computer is the quantum bit (qubit), the quantum entity that corresponds to the bit in a classical computer, but which can store a superposition of 0 and 1~\cite{Nielsen2000}.  The main challenge in building a quantum computer is that the qubit states are very fragile and susceptible to noise.  Surface codes~\cite{kitaev2003fault, dennis2002topological, fowler2012surface, terhal2015quantum} are two-dimensional structures of qubits located on a regular grid which provide fault tolerance by entangling the qubits. In the surface code, logical qubits are topologically protected, which means that only strings of bit flips that stretch from one side to the other of the code cause logical bit flips, whereas topologically trivial loops (contractable to a point) do not. In recent years, experiments have taken first steps in quantum error correction in several promising quantum-computing architectures, e.g., superconducting circuits~\cite{Reed2012, Shankar2013, Riste2015, Kelly2015, Corcoles2015, Ofek2016, Takita2017, Kockum2019a, Gong2019, KraglundAndersen2019}, trapped ions~\cite{Chiaverini2004, Schindler2011, Lanyon2013, Nigg2014, Linke2017}, and photonics~\cite{Yao2012, Bell2014}, and work continues towards large-scale implementation of surface codes.

Even though the surface-code architecture provides extra protection to logical qubits, the physical qubits are still susceptible to noise causing bit-flip or phase-flip errors. Such errors need to be monitored and corrected before they proliferate and create non-trivial strings that cause logical failure. The challenge with correcting quantum-mechanical errors is that the errors themselves cannot be detected (because such measurements would destroy the quantum superposition of states), but only the syndrome, corresponding in the surface codes to local 4-qubit parity measurements, can.  An algorithm that provides a set of recovery operations for correction of the error given a syndrome is called a \textit{decoder}. As the syndrome does not uniquely determine the errors, the decoder needs to incorporate the statistics of errors corresponding to any given syndrome. Optimal decoders, which give the highest theoretically possible error-correction success rate, are generally hard to find, except for the simplest hypothetical types of noise.

Many types of decoder algorithms exist that deal in different ways with the lack of uniqueness in the mapping from syndrome to error configuration. Methods range from Monte Carlo-based decoders~\cite{PhysRevLett.109.160503,hutter2014efficient},  cellular automata~\cite{herold2015cellular,kubica2018cellular}, renormalization group~\cite{duclos2010fast}, as well as various types of neural-network-based decoders~\cite{PhysRevLett.119.030501, krastanov2017deep, varsamopoulos2017decoding, baireuther2018machine, breuckmann2018scalable, chamberland2018deep, ni2018neural, sweke2018reinforcement, andreasson2018quantum, nautrup2018optimizing, maskara2019advantages,  chinni2019neural, colomer2019reinforcement}, which is also the tool used in the present paper.  The benchmark algorithm for the decoding problem is \ac{MWPM}~\cite{edmonds1965paths, fowler2015minimum, PhysRevA.90.032326}, which is a graph algorithm for pairwise matching of syndrome defects that is based on the assumption that the most likely error configuration is one that corresponds to the minimum number of errors. However, this does not take into account that different error channels may have different probabilities (biased noise), or that syndrome defects will in general be correlated.   

For a decoder to be used for actual operation in a quantum computer, not only correction success rate, but also speed, is a crucial factor. A long delay for calculating error correcting operations will not only slow down the calculations, but also make the code susceptible to additional errors. For this reason, decoders based on algorithms that do extensive sampling of the configuration space on the fly, such as Monte Carlo-based decoders~\cite{PhysRevLett.109.160503}, may not be viable as practical decoders. Instead, using some level of pre-training to generate and store information for fast retrieval will likely be necessary. Tabulating the information of syndrome versus most likely logical error is expected to be prohibitively expensive in terms of both storage and training, and slow to access, for anything but very small codes. Given these constraints, the need for pre-training, the massive state space and corresponding amount of data, it is natural to consider machine-learning solutions, especially given the recent deep-learning revolution~\cite{lecun2015deep, goodfellow2016deep} and its applications within quantum physics~\cite{carleo2017solving, carrasquilla2017machine, van2017learning}.

In this paper, we use deep reinforcement learning~\cite{mnih2013playing, mnih2015human}, expanding on the framework for error correction in the toric code (i.e., surface code with periodic boundary conditions) introduced by \citet{andreasson2018quantum}. Reinforcement learning and \ac{DRL} has recently emerged as a promising tool for various quantum control tasks~\cite{PhysRevX.8.031086, PhysRevX.8.031084}. 
In Ref.~\cite{andreasson2018quantum}, only uncorrelated noise (with independent bit- and phase-flip errors) was considered and it was found that the DRL decoder could achieve success rates of error correction on par with MWPM. In the present work, we consider depolarizing noise ($p_x = p_y = p_z$) and find that a similar decoder can outperform MWPM for moderate code size $d\leq 9$. The decoder trained on depolarizing noise is also found to be quite versatile, having MWPM success rates on uncorrelated noise, as well as giving intermediate performance on biased noise. Similarly to the previous work we do not consider syndrome measurement errors, but focus on mastering the more elementary but nevertheless challenging task of efficiently decoding a perfect syndrome with depolarizing noise.    

A decoder based on \ac{DRL} has the potential to offer an ideal balance between calculations on the fly and pre-training. The information about the proper error correction string for a given syndrome is stored in a very efficient way, using two principles: 
\begin{enumerate}[1)]
\item The step-by-step decoding using the pre-trained neural network generates an effective tree structure where many different syndromes will reduce to the same syndrome after one operation, such that subsequent correction steps will use the same information, iteratively reducing the complexity.
\item The deep neural network is a `generalizer' which can spot and draw conclusions from common features of different syndromes, including syndromes that have not been seen during training.
\end{enumerate}

The paper is organized as follows. In Sec.~\ref{sec:toric}, we give a brief introduction to quantum error correction for the toric code. In Sec.~\ref{sec:DRL}, we introduce deep reinforcement learning and Q-learning, and discuss how these are implemented in training and utilizing the decoder. In Sec.~\ref{sec:results}, the performance of the DRL decoder is presented and benchmarked against both MWPM and analytic expression valid for low error rates. We summarize the main results and give an outlook to further developments in Sec.~\ref{sec:conclusion}. 


\section{Toric code}
\label{sec:toric}

\begin{figure}
\centering
\includegraphics[width=\linewidth]{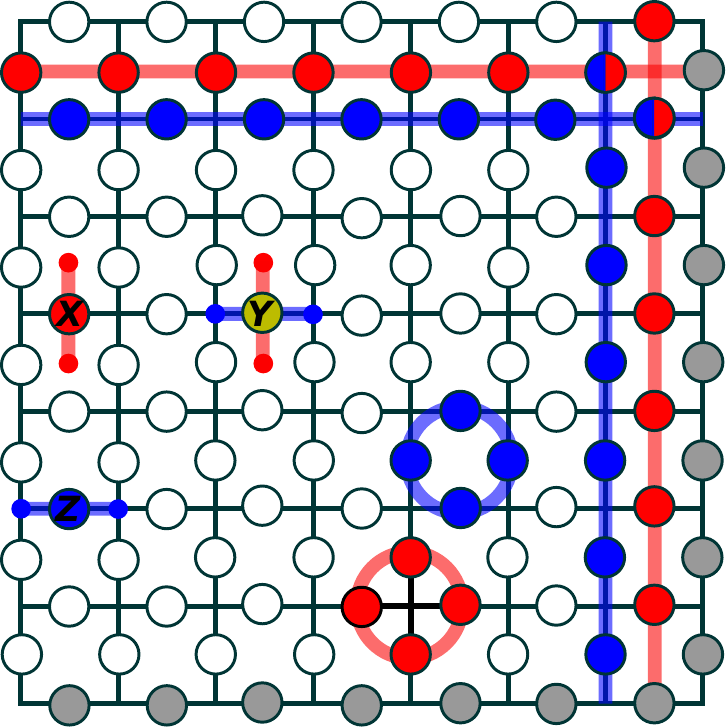}
\caption{A $d=9$ toric code showing the basic operations. Circles represent physical qubits, with shading showing periodic boundaries. Bit flip $X$ (red), phase flip $Z$ (blue), and $Y \sim X Z$ (yellow) errors with corresponding plaquette and vertex ``defects" as end points of error chains. The defects are measured by the plaquette ($\otimes Z$) and vertex ($\otimes X$) parity-check operators, respectively. Also shown are logical bit and phase flip operators corresponding to closed loops spanning the torus.
\label{fig:syndrome_0}}
\end{figure}

The toric code in the form considered here consists of a two-dimensional quadratic grid of physical qubits with periodic boundary conditions. In this section, we provide a high-level summary of the main concepts relevant for our study and refer the reader to the literature for more details~\cite{kitaev2003fault, dennis2002topological, fowler2012surface, terhal2015quantum}.  
A $d\times d$ grid contains $2d^2$ qubits corresponding to a Hilbert space of $2^{2d^2}$ states, out of which four will form the logical code space. That is, it encodes a 4-fold qudit corresponding to two qubits, which we will nevertheless refer to as the logical qubit. 
It is a stabilizer code where a large set of commuting local parity check operators (the stabilizers) split the state space into distinct sectors. 

The stabilizers for the toric code are divided into two types, here represented as plaquette and vertex operators, consisting of products of Pauli $Z$ or $X$ operators on the four qubits on a plaquette or vertex (see Fig.~\ref{fig:syndrome_0}), respectively. Eigenstates of the full set of stabilizers, with eigenvalue $\pm 1$ on each plaquette and vertex of the lattice, are globally entangled, which provides the basic robustness to errors. The logical qubit corresponds to the sector with eigenvalue $+1$ on all stabilizers. We will refer to a stabilizer with eigenvalue $-1$ as a plaquette or vertex defect. A single bit flip $X$ or phase flip $Z$ on a state in the qubit sector will produce a pair of defects on neighboring plaquettes or vertices, with Pauli $Y \sim X Z$ giving both pairs of defects, as shown in Fig.~\ref{fig:syndrome_0}.

The set of stabilizer defects corresponding to any given configuration of $X$, $Y$, or $Z$ operations on a state in the logical sector is called the syndrome. Logical operations, which map between the different states in the logical sector, are given by strings of $X$ or $Z$ operators that encircle the torus, corresponding to logical bit-flip and phase-flip operations, respectively (see Fig.~\ref{fig:syndrome_0}). The shortest loop that can encircle the torus has length $d$; correspondingly, the \textit{code distance} is $d$. For simplicity, we consider only odd $d$, as there is an odd-even effect in some quantitative aspects of the problem. The toric code is an example of a topological code, as the logical operations correspond to `non-contractible' loops on the torus, whereas products of stabilizers can only generate `contractible' loops.  

\begin{figure}
\includegraphics[width=\linewidth]{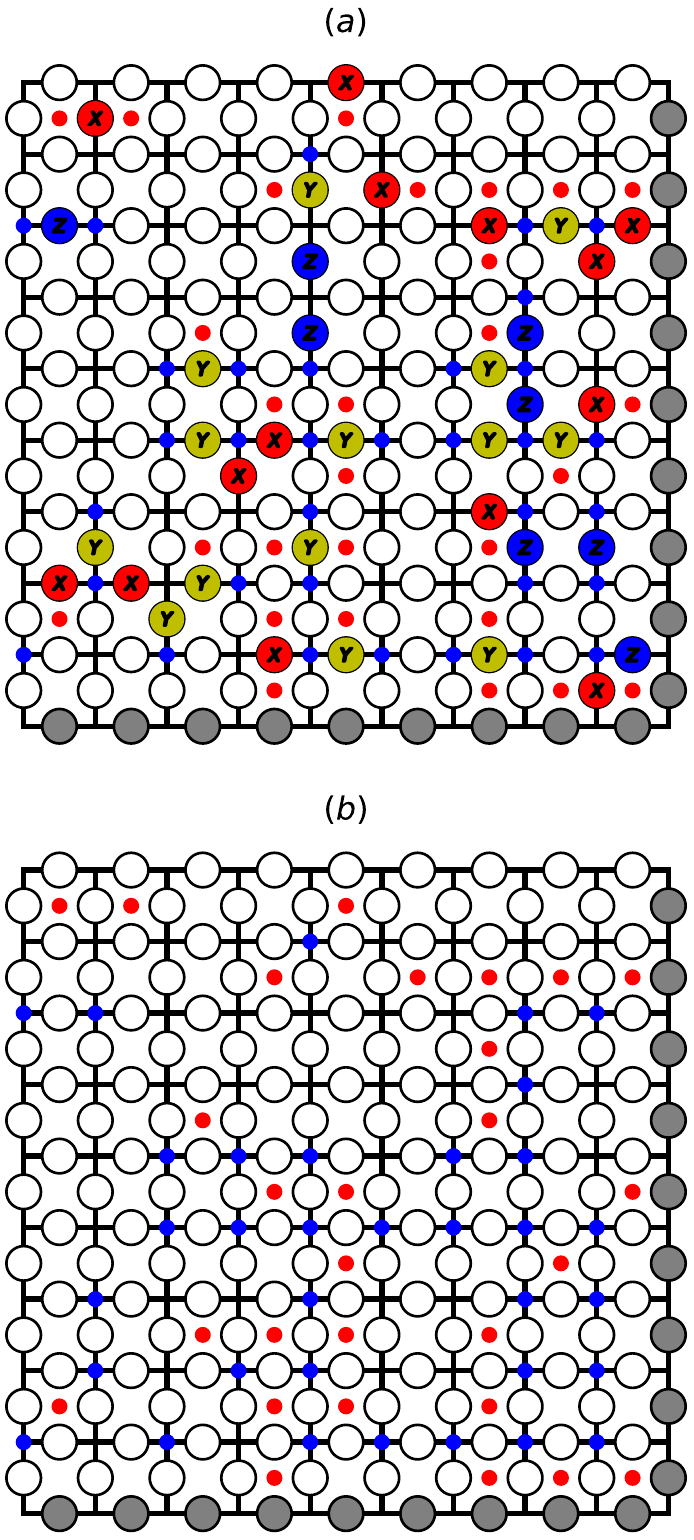}
\caption{Example of a random configuration of qubit errors on a $d=9$ toric code. (a) The qubit state and the corresponding syndrome forming an error chain. (b) Syndrome given by plaquette and vertex defects. The objective of the \ac{DRL} decoder is to find a correction string which is consistent with the syndrome and which takes the minimal number of qubit operations. The benchmark \ac{MWPM} decoder instead treats the plaquette and vertex configurations as separate graph problems, suggesting the shortest independent correction chains of $X$ and $Z$. The full decoding sequence for this syndrome using the DRL decoder is shown at \href{https://github.com/mats-granath/toric-RL-decoder}{github.com/mats-granath/toric-RL-decoder}.
\label{fig:syndrome}}
\end{figure}

Figure~\ref{fig:syndrome}(a) shows an example of an error configuration (also referred to as an error chain) on a $d=9$ toric code together with the corresponding syndrome, generated randomly at an error rate $p = 0.22$. Visible for the decoder is only the syndrome [Fig.~\ref{fig:syndrome}(b)] based upon which the decoder should suggest a sequence of operations (a correction chain) that eliminates the syndrome in such a way that it is least likely to cause a logical bit- and/or phase-flip operation. To evaluate the success rate of a correction chain for a given syndrome, it should be complemented by the full distribution of error chains corresponding to that syndrome, to calculate which fraction of error+correction chains contain an odd number of logical operations of any type.


\section{Deep reinforcement learning algorithm}
\label{sec:DRL}

The \ac{DRL}-based decoder presented in this paper is an agent utilizing reinforcement learning together with a deep convolutional neural network, called the Q-network, for approximation of Q-values. The agent suggests, step by step, a sequence of corrections that eliminates all defects in the system as illustrated in Fig.~\ref{fig:Q-example} (see also Figs.~\ref{fig:sequence_correction_fail} and \ref{fig:sequence_correction_success} in Appendix~\ref{appendix:selected_episodes}).

\subsection{Q-learning}

\begin{figure}
\centering
\includegraphics[width=\linewidth]{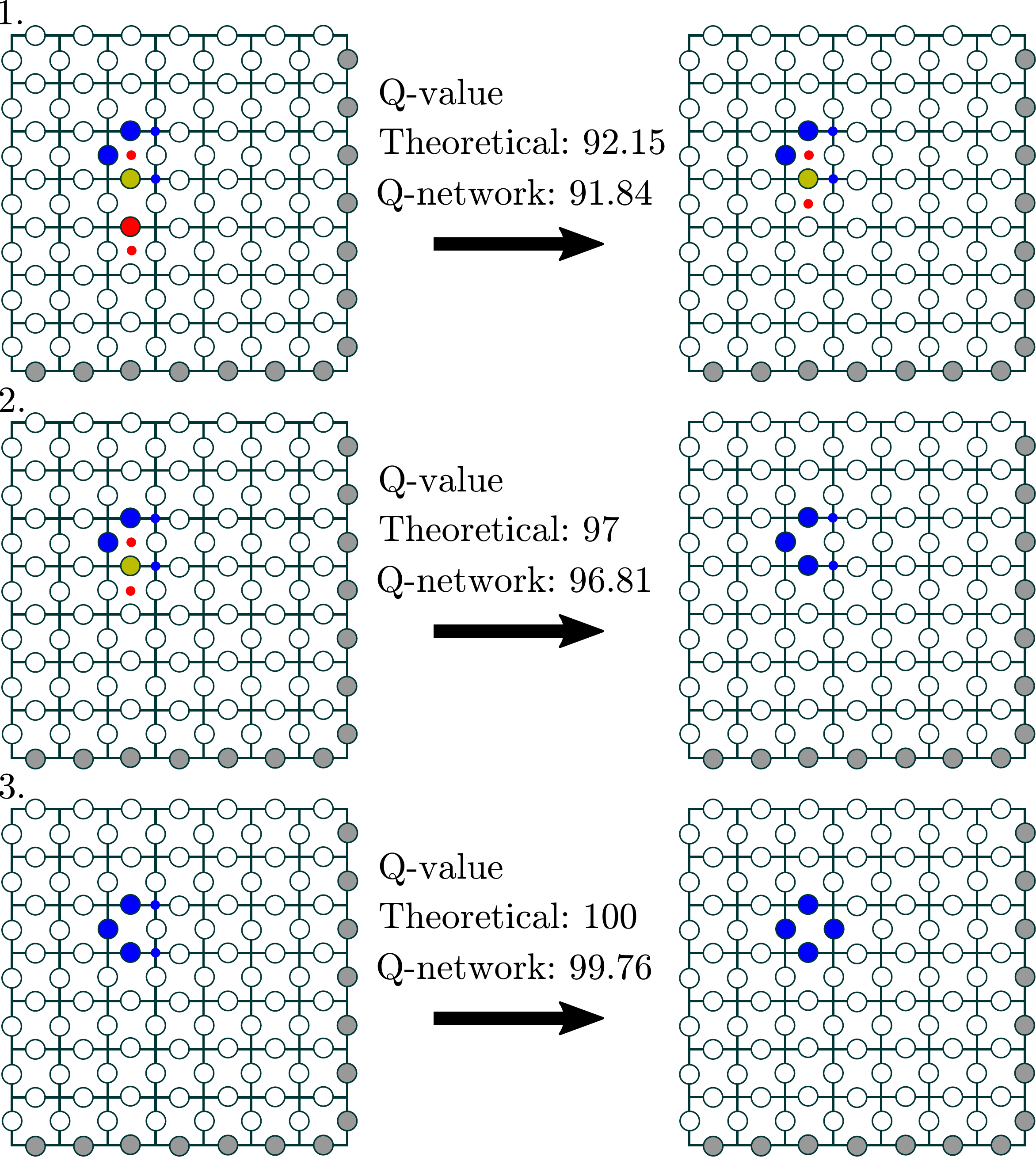}
\caption{Value functions $V(s) = \max_a Q(s,a)$ for a sequence of syndromes corresponding to a particular error chain, using the reward scheme in Eq.~(\ref{equ:reward_definition}) with $\gamma = 0.95$. For this simple syndrome, the optimal sequence is three steps long and the theoretical state values are compared to those output by the Q-network. The error chain itself is irrelevant to the correction sequence; only the syndrome is important.}
\label{fig:Q-example}
\end{figure}

The purpose of Q-learning~\cite{sutton2018reinforcement} is for an agent to learn a policy, $\pi(s,a)$, that prescribes what action $a$ to take in state $s$. An optimal policy maximizes the future cumulative reward of actions within a Markov decision process with the rewards provided by the environment, depending on the initial and final states and the action $r_a (s,s')$. In this paper, we use a deterministic reward scheme, as discussed below.  To measure the future cumulative reward, the action value function, or Q-function, is given by
\begin{equation}
Q_{\pi}(s_t, a_t) = \mathbf{E}_{\pi} \mleft[ r_t + \gamma r_{t+1} + \gamma^2 r_{t+2} + \ldots \mright],
\end{equation} 
where action $a_t$ is taken at time $t$, and subsequently following the policy $\pi$, with $\gamma \leq 1$ a discounting factor. The Q-function corresponding to the optimal policy satisfies the Bellman equation
\begin{equation}
Q(s_t, a_t) = r + \gamma \max_{a'}Q(s_{t+1}, a'),
\end{equation}
such that the optimal policy will self-consistently correspond to the action maximizing $Q$. As discussed in more detail in Sec.~\ref{sec:training}, we use one-step Q-learning, in which the current measure of $Q(s,a)$ is updated by explicit use of the Bellman equation with some learning rate $\alpha$, using $\epsilon$-greedy exploration. 

The reward scheme that we use is given by 
\begin{equation}
r_t =  
    \begin{cases} 
    100 & \mbox{if episode terminates at step $t+1$}\\
    E_{t} - E_{t+1} &\mbox{otherwise},
    \end{cases}
    \label{equ:reward_definition}
\end{equation}
where $E_t$ represents the number of defects in the syndrome at step $t$, such that $X$ and $Z$ operators can give reward $-2$, $0$, or $2$, whereas $Y$ operators can give reward $-4$, $-2$, $0$ , $2$, or $4$. The terminal reward, given a discounting factor $\gamma<1$, incites the agent to correct the full syndrome in the minimal number of steps. The explicit reward for eliminating defects is implemented to speed up convergence, without which the agent would have to find terminal states by completely random exploration.  The reward scheme is not expected to give an optimally performing decoder~\cite{andreasson2018quantum, colomer2019reinforcement}; rather than using the statistics of error chains in an unbiased fashion, it makes the assumption that the most likely error chain is the shortest. As expected (see Sec.~\ref{sec:results}), for biased noise this gives sub-optimal performance. 

Figure \ref{fig:Q-example} shows an example of Q-network estimated and exact state values $V(s) = \max_a Q(s,a)$ for an example syndrome, showing that the Q-network gives a quantitatively accurate representation of Q-values. The numerical accuracy in general deteriorates the larger the syndrome is, i.e., the further it is removed from the terminal state.  

\subsubsection{Efficient Q-network representation}

\begin{figure}
\centering
\includegraphics[width=\linewidth]{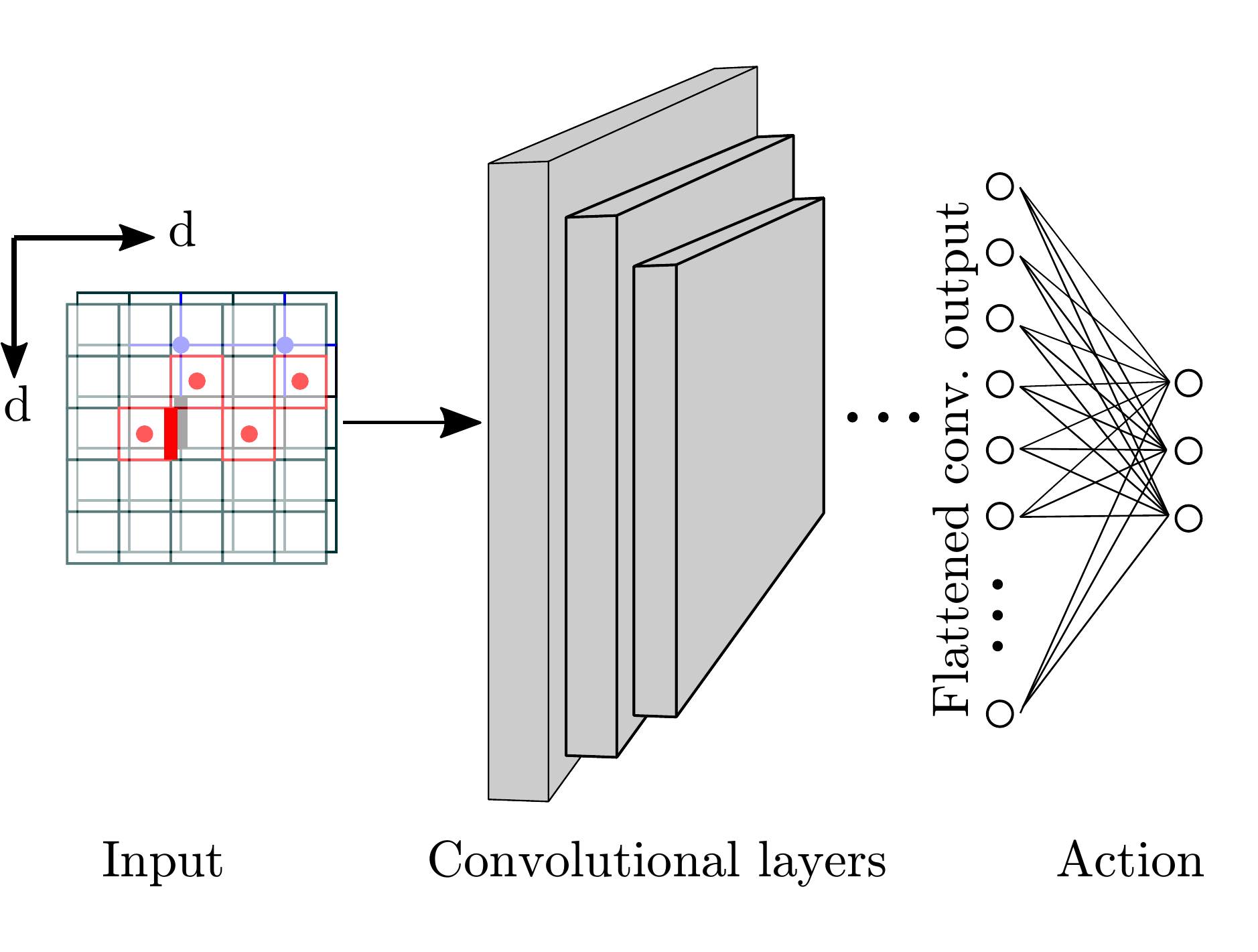}
\caption{Input-output structure of the deep Q-network. The input is a perspective, $P$, constructed from the syndrome, $s$, as shown in Fig.\ \ref{fig:spo}. The hidden layers consist primarily of convolutional layers (see Appendix \ref{appendix:model_definition_hyperparameters} for details). The output is the three action Q-values, $Q(P, a, \theta)$, for $a \in \{X, Y, Z\}$ operators on the marked (bold) qubit, with $\theta$ representing the current state of the network.
\label{fig:network}}
\end{figure}

To improve the representational capacity of the Q-network, we use an efficient state-action space representation, which was suggested in Ref.~\cite{andreasson2018quantum} for bit-flip operations and which we now extend to general $X$, $Y$, and $Z$ operations. It is built on three basic concepts: 
\begin{itemize}
\item By having the Q-network only output action values for one particular qubit, the representational complexity can be reduced significantly.
\item Due to the periodic boundary conditions of the toric code, only the relative positions of syndrome defects are important, i.e., arbitrary translations and four-fold rotations are allowed.
\item The converged decoder will never operate on a qubit which is not adjacent to any syndrome defect. Consequently, we have no need to calculate Q-values for such actions.
\end{itemize}

\begin{figure}
\centering
\includegraphics[width=\linewidth]{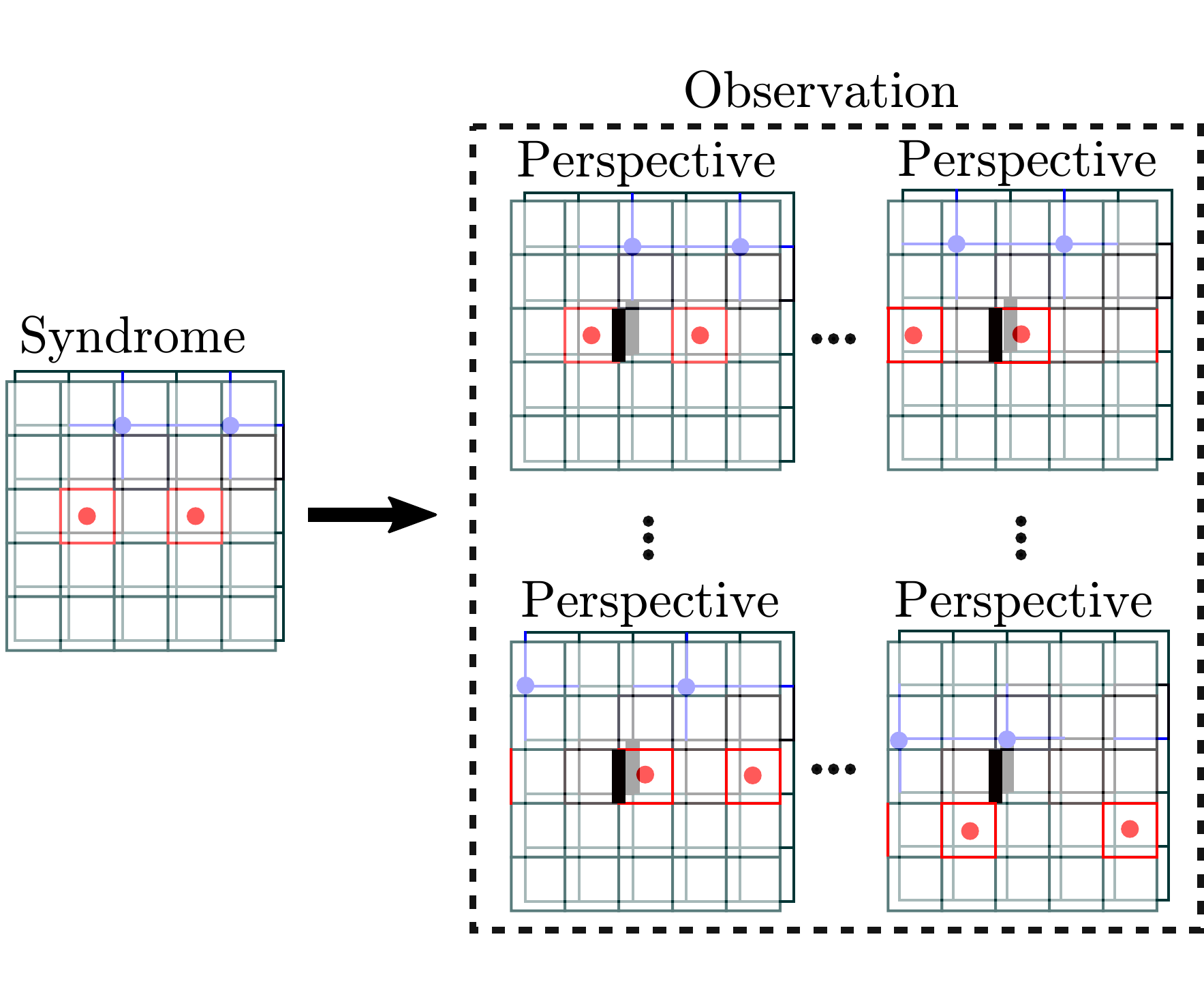}
\caption{Expanded representation of a syndrome into different perspectives, based on rotations and translations, used for compact processing in the Q-network (Fig.~\ref{fig:network}). Only the syndrome, visible to the network, is shown, not the physical qubits. The two-layer structure corresponds to seperate channels of input for vertex and plaquette defects. The set of all perspectives form an observation, $O = \{P_1, P_2, ..., P_{N_{per}}\}$.
\label{fig:spo}}
\end{figure}

The Q-network takes input in the form of two channels of $d \times d$ matrices, corresponding to the location of vertex and plaquette defects, respectively. The output is the three Q-values for $X$, $Y$, and $Z$ operations on one particular qubit, in a fixed location $\vec{r}_0$ with respect to an external reference frame, as indicated in Fig.~\ref{fig:network}. To obtain the full set of action values for a syndrome, we thus successively translate and rotate the syndrome to locate each qubit at location $\vec{r}_0$. Each such matrix representation of the syndrome, with a particular qubit at $\vec{r}_0$, is called a "perspective", and the whole set of perspectives makes up an "observation", as exemplified in Fig.~\ref{fig:spo}. In the observation, we only include perspectives for qubits that are adjacent to a syndrome defect. 

\begin{figure}
\centering
\includegraphics[width=\linewidth]{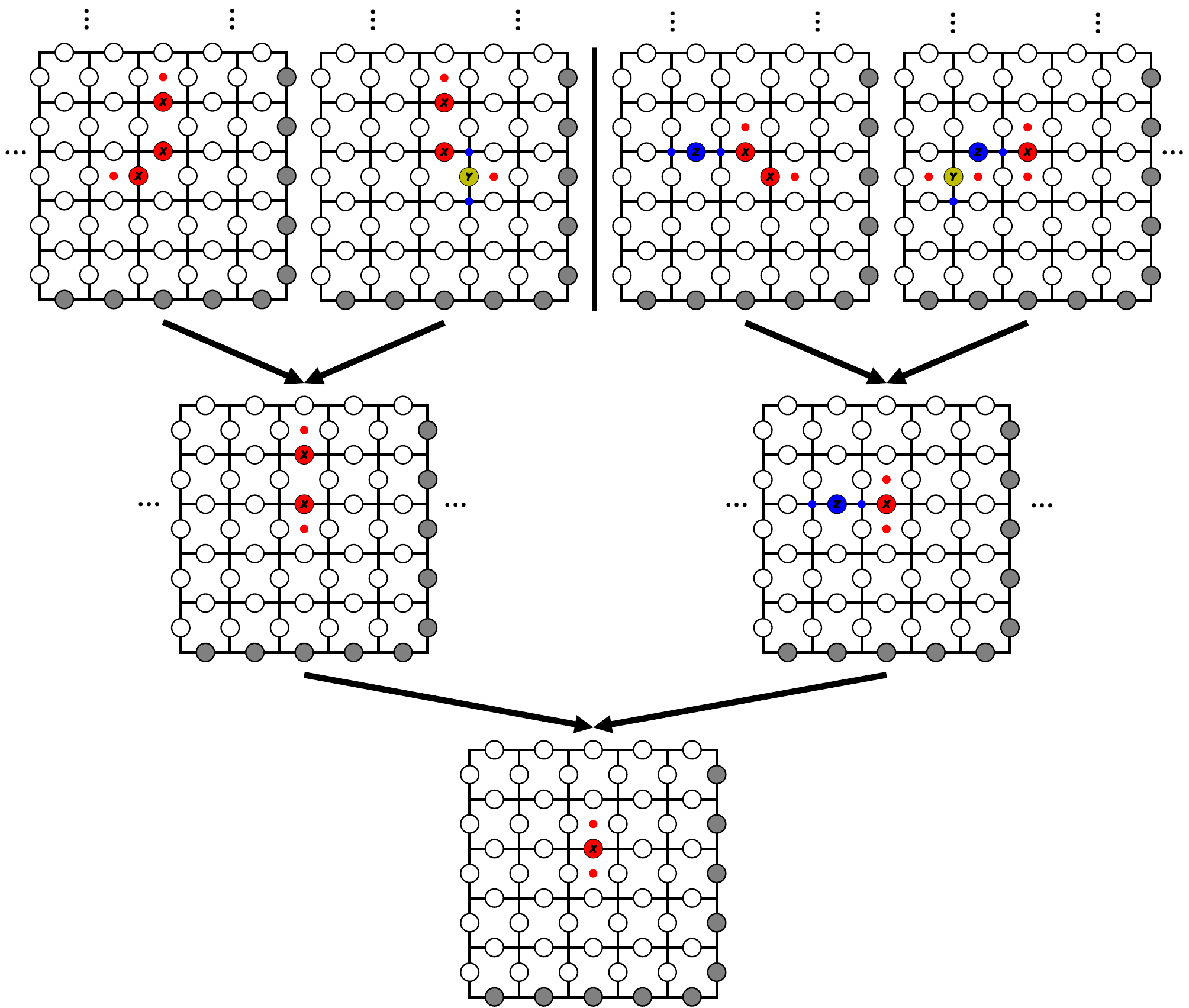}
\caption{Schematic of the operation of the \ac{DRL} decoder for several syndromes that successively reduce to a smaller subset of syndromes through step-by-step decoding. Top left are two syndromes that after one step of decoding reduces to the same syndrome, and similarly to the right. Both these branches in turn reduce to the same syndrome after the next decoding step. In this way, the complexity of the decoding problem is reduced, compared to decoding each high-level syndrome independently.
\label{fig:tree}}
\end{figure}

To obtain the full relevant Q-function of a syndrome, the Q-function of each individual perspective of an observation is calculated. In decoding mode, the agent chooses greedily the action with the highest Q-value. After the chosen action has been performed, a new syndrome is produced and the process repeats until no defects remain. As discussed in the introduction, and exemplified in Fig.~\ref{fig:tree}, the DRL decoding framework gives a compact structure for information storage and utilization: using a neural network  to generalize information between syndromes and using step-by-step decoding to successively reduce syndromes to a smaller subset.   

\subsection{Training the Q-network}
\label{sec:training}

The neural network is trained using the deep Q-learning algorithm utilizing prioritized experience replay~\cite{mnih2015human, schaul2015prioritized}. 
To increase stability, two architecturally equivalent neural networks are used, the regular Q-network, with parameters $\theta$, and the target Q-network, with parameters $\theta_T$. The target network is synchronized with the Q-network on a set interval.

Experience replay saves every transition in a memory buffer, from which the agent randomly samples a mini-batch of transitions used to update the Q-network. Instead of sampling the mini-batch uniformly, as is done with regular experience replay, prioritized experience replay prioritizes importance when sampling. This importance is measured with the absolute value of the temporal difference (TD) error,
\begin{equation}
\delta_j = r_j + \gamma \max_{a}(Q(s'_{j}, a;\theta_T)) - Q(s_j, a_j;\theta)\,,
\end{equation}
where the state$/$syndrome $s'_{j}$ follows from action $a_j$ on state$/$syndrome $s_j$, and where the expression $Q(s, a;\theta)$ implies choosing the appropriate perspective for the Q-network that corresponds to action $a$ in syndrome $s$.

Following Ref.~\cite{schaul2015prioritized}, the probability of sampling a transition $j$ from the memory buffer is given by
$P_j = \abs{\delta_j}^{\alpha}/\sum_k \abs{\delta_k}^{\alpha}$
such that values with higher TD-error are more likely to be sampled. Here, $\alpha$ controls the amount of prioritization used ($\alpha = 0$ corresponding to uniform sampling) and $k=1,...,M$, with $M$ the size of the memory buffer. Using non-uniform sampling in this way, however, skews the learning away from the probability distribution used to generate experiences. To partially compensate for this, importance-sampling weights are introduced according to 
$w_j = (M \cdot P_j)^{-\beta}$,
with the product of the weights and TD-error, $w_j \cdot \delta_j$, used as the loss during stochastic gradient descent training of the network. Here $\beta$ controls the extent of compensation of the prioritized sampling, with $\beta = 1$ corresponding to full compensation.

The training can be divided into two stages: the action stage and the learning stage. Pseudo-code of the algorithm used for training is shown in Algorithm~\ref{alg_RL_decoder}. The training starts with the action stage. Given a syndrome $s_t$, the agent suggests an action $a_t$ following an $\epsilon$-greedy policy, such that with probability ($1-\epsilon$) the agent takes the action with the highest Q-value; otherwise a random action is followed. The agent receives a reward, $r_t$, and the syndrome, $s'_t=s_{t+1}$, that follows from the action $a_t$. The transition is stored as a tuple, $T = (P_t, a_t,r_t,s_{t+1}, \Theta_{t+1})$, where $\Theta_{t+1}$ is a Boolean containing the information whether $s_{t+1}$ is a terminal state (there are no defects left) or not. 

\begin{algorithm}
\SetAlgoNoLine
 \While{defects remain}{
  Get observation $O_t$ corresponding to syndrome $s_t$.\;
  With probability $\epsilon$ select random action $a_t$ and corresponding perspective $P_t$.\;
  Otherwise select: \newline 
  $\{P_t,a_t\} = \text{argmax}_{P,a}(Q(P, a; \theta)_{P \in O_t}$.\;
  Execute action $a_t$ and observe reward $r_t$ and syndrome $s_{t+1}$.\;
  Store transition ($P_t, a_t, r_t, s_{t+1}, \Theta_{t+1}$) in replay memory.\;
  Sample random mini-batch of transitions, $\{T_j\}_{j=1}^N$, from replay memory using prioritized sampling.\;
  Calculate weights used for weighted importance sampling $w_j$.\;
  If terminal state reached, set $y_j = r_j$; otherwise, set $y_j = r_j + \gamma \max_{a}Q(s'_j, a; \theta_T)$.\;
  Perform gradient descent step on $w_j \cdot \abs{y_j - Q(P_j, a_j;\theta)}$ with respect to the network parameter $\theta$.\;
  Every C steps synchronize the target network with the policy network, $\theta_T = \theta$.\;
 }
 \caption{Training the \ac{DRL} agent decoder}
 \label{alg_RL_decoder}
\end{algorithm}

After the action stage, the agent continues with the learning stage. For that we use \ac{SGD} and the tuples stored in the replay memory. A mini-batch of $N$ transitions, $\{T_j=(P_j,a_j,r_j,s'_j,\Theta_j)\}_{j=1}^N$, is sampled from the replay memory with replacement. The training target value for the policy Q-network is given by $y_j = r_j$ if $\Theta_j=1$, and $y_j = r_j + \gamma \max_{a}Q(s'_{j}, a; \theta_t)$ otherwise. 

The agents are initially trained with an error rate of $10\%$ and further during the training with syndromes up to $30\%$ error rate. Details of network architectures and hyperparameters are found in Appendix~\ref{appendix:model_definition_hyperparameters}.


\section{Results}
\label{sec:results}

\subsection{Depolarizing noise}

\begin{figure}
\centering
\includegraphics[width=\linewidth]{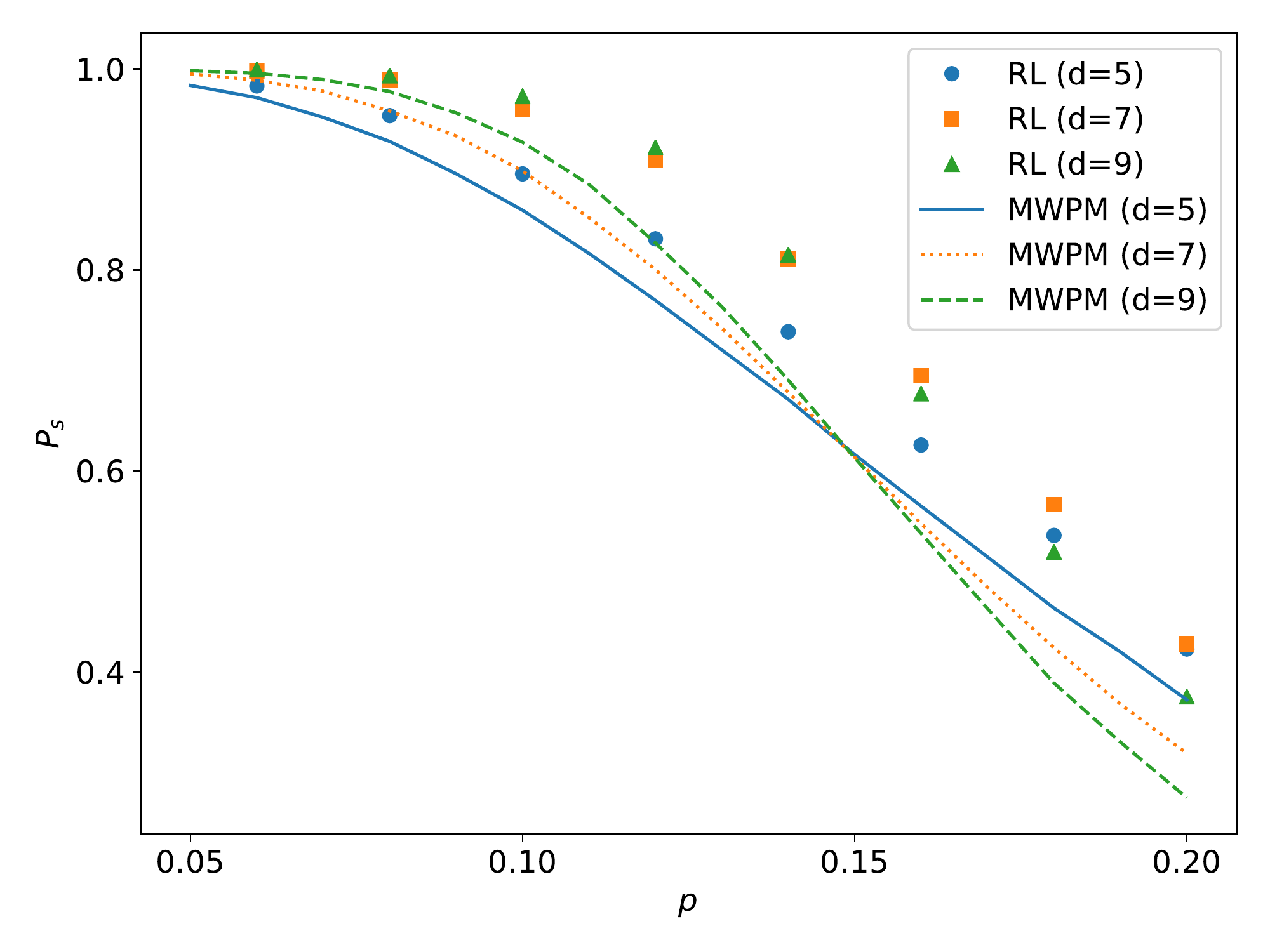}
\caption{Error correction success rate, $P_s$, for the DRL decoder on depolarizing noise, as a function of total error probability $p$, for system sizes $d = 5, 7, 9$ (blue circles, orange squares, and green triangles, respectively), and compared to the corresponding results using the MWPM algorithm (blue solid curve, orange dotted curve, and dashed green curve, respectively). The \ac{DRL}-based algorithm outperforms the MWPM-based algorithm for all these system sizes and error rates.
\label{fig:high_p_plot}}
\end{figure}

\begin{figure}
\centering
\includegraphics[width=\linewidth]{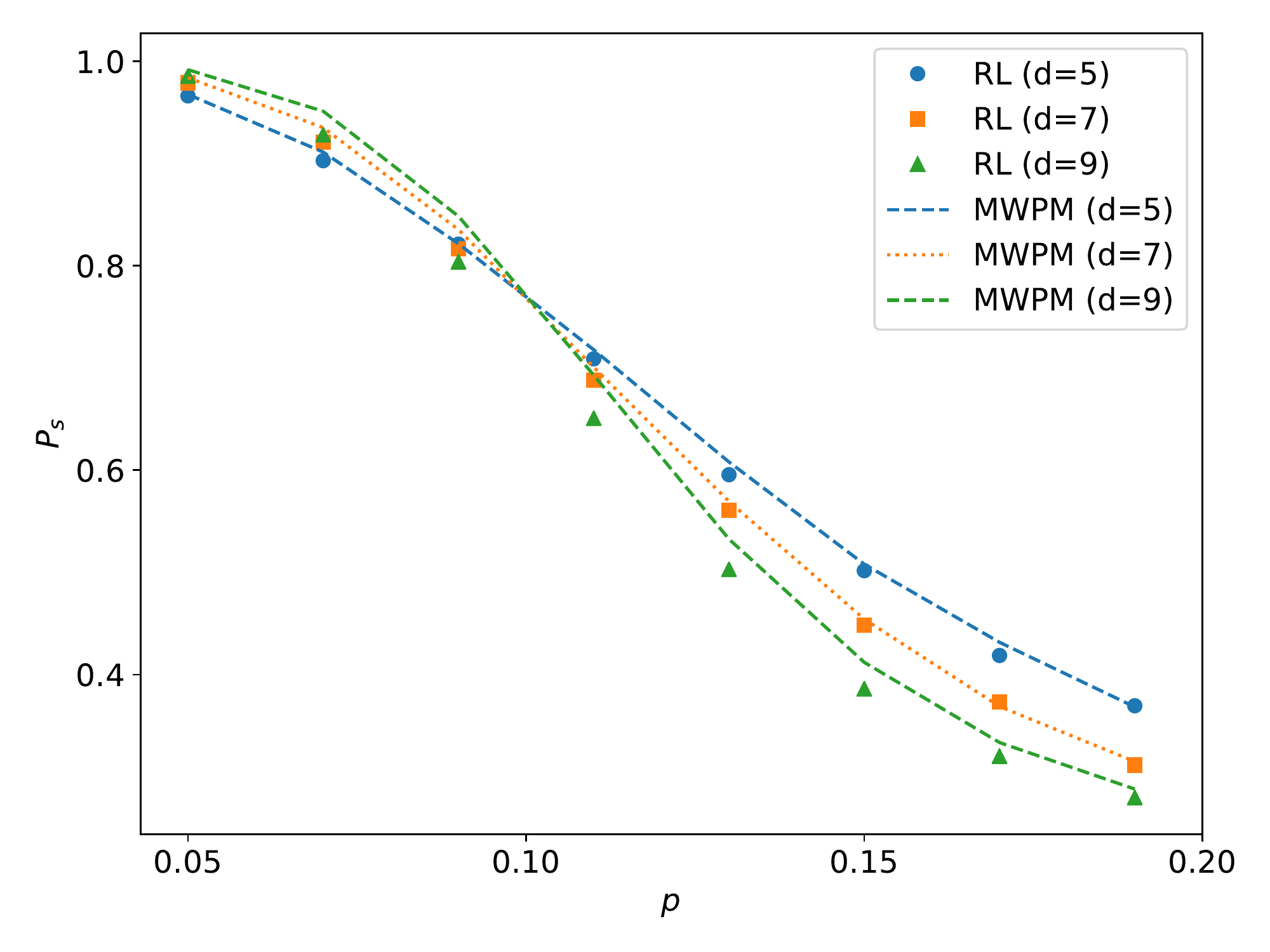}
\caption{Error correction success rate, $P_s$, for the DRL decoder trained on depolarizing noise, when applied to pure bit-flip noise, as a function of error probability $p$. Dashed curves show the corresponding results using the MWPM algorithm. 
\label{fig:high_p_plot_only_x}}
\end{figure}

The main result of the paper is displayed in Fig.~\ref{fig:high_p_plot}, where the error-correction success rate for depolarizing noise, $p_x = p_y = p_z = p / 3$, is shown for decoders trained at three different code dimensions. This is compared to MWPM, which treats the plaquette and vertex defects as separate graph problems. See comment
\footnote{The MWPM decoder assumes that $X$ and $Z$ errors are uncorrelated, with independent error rates $p_x = p_z = 2p/3$ and correspondingly $p_y = (2p/3)^2$. The MWPM success rate for that problem would be $P_{S}(p) = (P_{S,X}(2p/3))^2$, with $P_{S,X}(p)$ corresponding to pure bit-flip noise (Fig.~\ref{fig:high_p_plot_only_x}). This expression is a good approximation to the MWPM success rate for depolarizing noise which is exact in the low-$p$ limit (see Appendix~\ref{appendix:small_error_rates}).}
for a discussion about the MWPM decoder for depolarizing noise. We thus find that the DRL decoder has a significantly higher error-correction success rate, which is achievable by learning to account for the correlations between plaquette and vertex defects.

From the crossing of the $d = 5$ and $d = 7$ error-correction success rates, we can identify a threshold of around $16.5$\% (for MWPM, the crossing is close to $15$\%), below which error correction can be guaranteed, were we able to increase $d$ arbitrarily. The deduced threshold is significantly below the theoretical limit of $18.9$\%~\cite{bombin2012strong, PhysRevLett.109.160503}, but, as discussed in the introduction, for a practical decoder this may not be the most important measure. We anticipate that the success rate and threshold can be enhanced by further developing the reward scheme to be based on success rate rather than minimum number of operations (work along these lines was recently presented by \citet{colomer2019reinforcement}). 

We also note that even though the $d = 9$ DRL decoder gives a significant improvement over MWPM, it has not fully converged to the optimal performance within the limitations of the algorithm, as indicated by the earlier crossing with $d=5$ and $d=7$.  We do not anticipate that this is a fundamental limitation of the DRL type decoder, but could be improved by a more efficient training scheme.

In Fig.~\ref{fig:high_p_plot_only_x}, we have employed the same DRL decoders, pre-trained on depolarizing noise, to decode pure bit-flip noise. Here, we find a performance for $d=5$ and $d=7$ which is very close to MWPM, thus reproducing the results of our first-generation DRL decoder from Ref.~\cite{andreasson2018quantum}. For $d=9$, the decoder has slightly worse performance, confirming that this decoder has not yet converged to optimal algorithmic performance.

\subsection{Asymptotic fail rates}

In addition to the MWPM benchmark, we also benchmark the DRL decoders for small error rates $p \xrightarrow{} 0$, by deriving analytical expressions (see Appendix~\ref{appendix:small_error_rates}) for the fail rate for depolarizing noise to lowest non-vanishing order in $p$. We can derive such fail rates for both the MWPM algorithm and the algorithm based on finding the shortest correction strings. The latter is similar to, but not exactly equivalent with, what we expect for the DRL decoder based on our reward scheme. These algorithms both have a fail rate that scales as $P_L\sim p^{\lceil \frac{d}{2}\rceil}$, but with different prefactors.

\begin{figure}
\centering
\includegraphics[width=\linewidth]{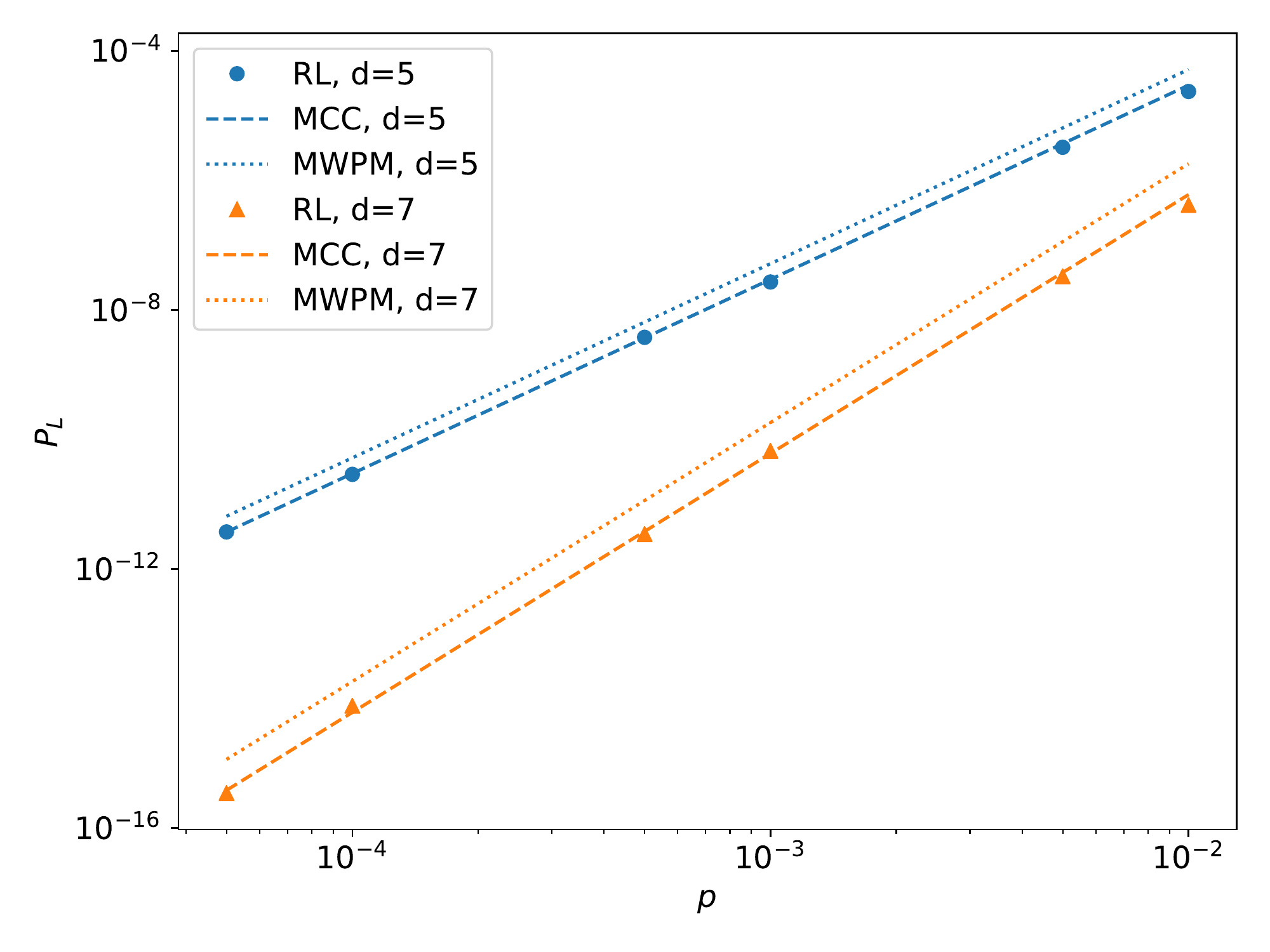}
\caption{Error correction fail rate, $P_L$, of the DRL decoder for depolarizing noise at low error rates. The dashed and dotted lines correspond to analytic expressions (see Appendix \ref{appendix:small_error_rates}), valid to lowest order in $p$, for a decoder that operates based on the minimal correction chain (MCC) or the MWPM algorithm.
\label{fig:low_p_plot}}
\end{figure}

In Fig.~\ref{fig:low_p_plot}, we confirm that the DRL decoder indeed performs ideally for $d = 5$ and $d = 7$ for short error chains, following very closely the algorithm based on minimal $X,Y,Z$ chains. Because of the excessive time consumption to generate good statistics for $d=9$, we have only compared the performance in the true asymptotic limit, i.e., the rate for only the shortest fallible error chains, as shown in Table~\ref{tab:p_l}, again confirming the sub-optimal performance for $d = 9$. In this limit, data is generated by only considering the sub-group of error chains that are in a single row or column, in contrast to generating completely random error chains that will very rarely fail.

\begin{table}
	\centering
	\caption{Comparison of  asymptotic logical fail rates $P_L$.}
	\label{tab:p_l}
	\begin{tabular}{c c c c}
		\hline
		& Analytic  & DRL decoder \\
		\hline
		$d = 5$ & 1.51e-3 & 1.45e-3\\
		$d = 7$ & 2.12e-5 & 2.07e-5\\
		$d = 9$ & 2.50e-7 & 4.30e-7\\
		\hline
	\end{tabular}
\end{table}


\subsection{Biased noise}

For the prospect of an operational decoder on a physical quantum computer, the noise is expected to be biased, such that phase-flip errors are relatively less or more likely~\cite{Ghosh2012, Yan2016, Gu2017, Klimov2018, Burnett2019, Lu2019}. To identify the exact error distribution is a challenging problem in itself (see, e.g., Ref.~\cite{PhysRevResearch.1.033092}), and the degree of bias can fluctuate in time~\cite{Klimov2018, Burnett2019, Lu2019}, so a decoder that can adequately decode biased noise without retraining might be an alternative. To quantify the performance of the DRL decoder for biased noise, we consider the probability of an error of any type $p$, probability of phase-flip error $p_z=p_{rel}p$, and consequently $p_x=p_y=(1-p_{rel})p/2$. Thus for $p_{rel} = 1$ the syndromes contain only $Z$ errors, which corresponds to uncorrelated noise, whereas $p_{rel} = 1/3$ corresponds to depolarizing noise. 

\begin{figure}
\centering
\includegraphics[width=\linewidth]{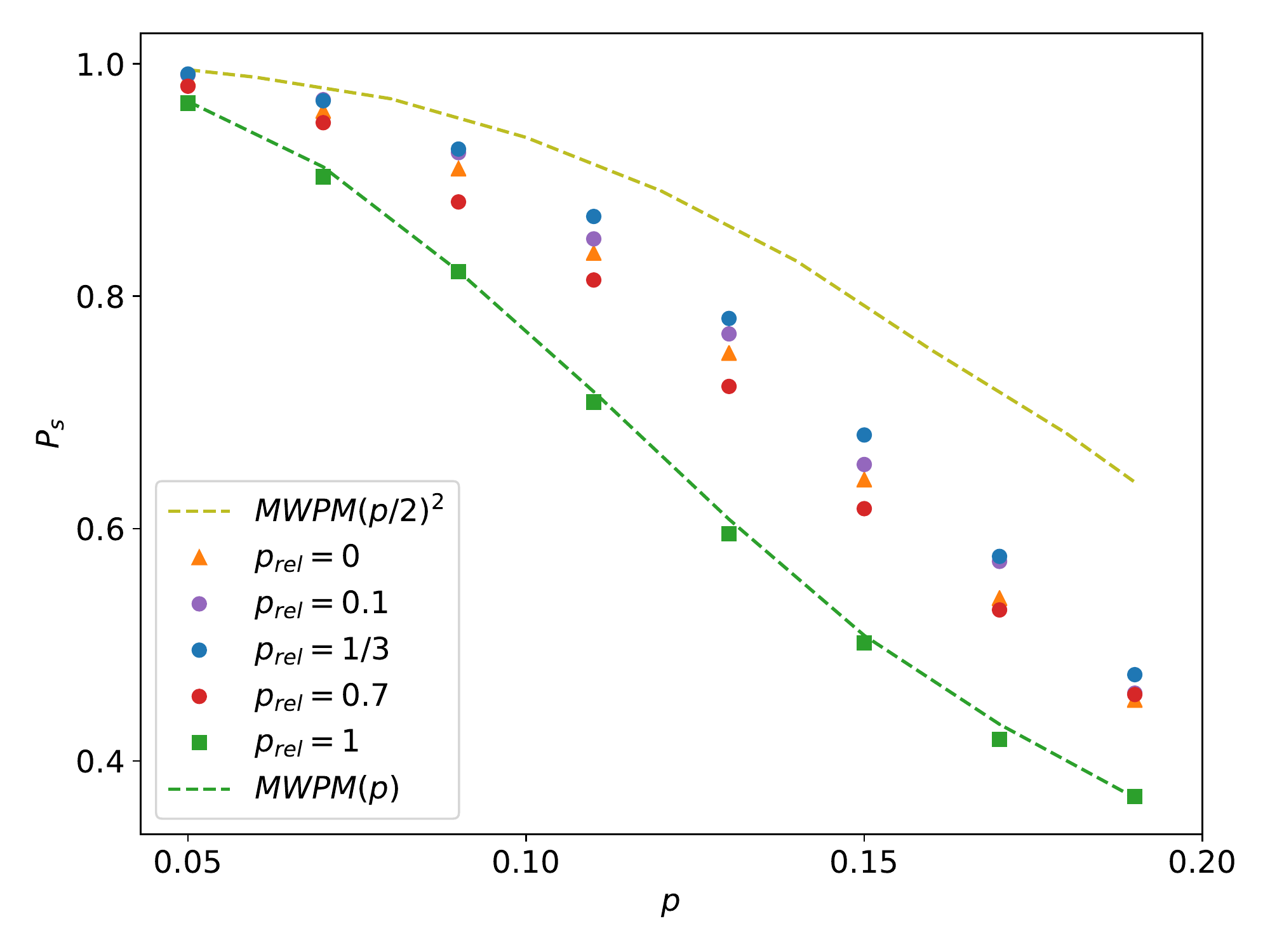}
\caption{Error-correction success rate for biased noise $p_z = p_{rel}p$, $p_x = p_y = (1-p_{rel})p/2$, using a decoder trained on depolarizing noise ($p_{rel} = 1/3$). For pure phase-flip noise ($p_{rel} = 1$), the decoder is compared to MWPM. The line MWPM$(p/2)^2$ indicates expected performance for an MWPM decoder designed explicitly for $p_z = 0$ noise.
\label{fig:biased_syndrome}}
\end{figure}

In Fig.~\ref{fig:biased_syndrome}, we show the success rate for the decoder on biased noise. We find that the highest success rate is attained for depolarizing noise, which also is what the decoder is trained for. 
We can understand this as a consequence of the superlinear decline (for low $p$) in success rate with the number of defects, such that the majority species dominates the outcome. At $p_{rel} =1/3$ there is an equal mean number of vertex and plaquette defects, while away from this limit, the number of either one or the other grows.     
That the operation of the trained DRL decoder is sub-optimal is clear from the limit $p_{rel}=0$, corresponding to only $X$ and $Y$ errors, which should, in principle, be a simpler decoding problem, similar to uncorrelated noise with independent error rates $p/2$ \footnote{Even though the limit $p_z=0$ corresponds to a surplus of plaquette defects versus vertex defects, the decoding problem is, in principle, equivalent to the problem of {\em non-coinciding} $X$ and $Z$ errors with error rates $p_x=p_z=p/2$: the decoder could first decode the vertex defects using $Y$ operators, and subsequently decode the remaining plaquette defects using $X$. The corresponding uncorrelated problem (with non-zero coincidence probability $(p/2)^2$) would have MWPM success rate $P_{S}= (P_{S,X}(p/2))^2$, which we expect is still a good approximation (for small $p$) and also close to optimal for this weakly correlated noise.}. Nevertheless, the decoder gives fair performance for the full range of biased noise, which may be an advantage over having a decoder which is specialized to a particular, potentially unknown, bias. 

\section{Conclusion}
\label{sec:conclusion}

We have shown how deep reinforcement learning can be used for quantum error correction of depolarizing noise ($p_x = p_y = p_z$) in the toric code, with significantly improved performance compared to the standard \ac{MWPM} algorithm. The advantage is gained by learning to account for the correlations between the vertex and plaquette defects. 
The super-MWPM performance for depolarizing noise was achieved for system sizes up to $d = 9$, corresponding to 162 qubits. However, by applying the trained decoder to decode pure bit-flip noise, ideal performance was only found for $d < 9$. For biased noise ($p_z \neq p_x = p_y$), the decoder gives fair, but sub-optimal, success rates.   

Several improvements of the complete algorithm are being explored, or would be interesting to explore. This includes using distributed reinforcement learning~\cite{Horgan2018} to enable the agent to explore the state space more efficiently and speed up the training. Moreover, it could be worth investigating the possibility of transferring the domain-specific knowledge (transfer learning) obtained from small grid instances to comparably larger grid sizes~\cite{Zhuang2019}. To combine the Q-learning with an element of active near-term exploration, such as that used by AlphaGo Zero~\cite{silver2017mastering, dalgaard2019global} would also be an interesting approach to investigate.    

The reward scheme used in this work is based on the heuristic to minimize the length of correction chains. This is a fair assumption for depolarizing noise, where $X$, $Y$, and $Z$ errors are equally likely. For biased noise, with greater or smaller probability of phase flip errors, training the decoder based on this assumption gives sub-optimal performance. Instead, the reward needs to be more closely linked to the actual distribution of error chains and syndromes.

In addition to improving the prowess for the problem discussed in this paper, further developments of the DRL decoder should include addressing syndrome measurement errors and non-toric topological codes~\cite{sweke2018reinforcement}. Even though the DRL-type decoder presented in this paper and in Refs.~\cite{andreasson2018quantum, colomer2019reinforcement} is still limited in scope, we have shown that it can flexibly address various types of noise, and in some regimes give super-MWPM performance. In addition, the information gathered from exploration is stored and used in an efficient and generalizable way using a deep neural network and step-by-step error correction, limiting both the complexity of concurrent calculations and the need for massive information storage, which may be instrumental for future operational decoders.      

\begin{acknowledgments}

Computations were performed on the Vera cluster at Chalmers Centre for Computational Science and Engineering (C3SE). We acknowledge financial support from the Knut and Alice Wallenberg foundation.

\end{acknowledgments}

\bibliography{Bib}


\appendix

\section{Small error rate}
\label{appendix:small_error_rates}

It is possible to derive a theoretical expression for the logical fail rate, that becomes exact in the limit of low error probabilities, by considering only the shortest possible error strings that may lead to an error given the decoding scheme. Here we derive such expressions for depolarizing noise $p_x = p_y = p_z = \frac{p}{3}$ for an algorithm which is based on correction using the minimum number of correction steps, and for an algorithm which is based on using MWPM separately on the graphs given by plaquette and vertex errors. The former algorithm, which we refer to as ``minimal correction chain'' (MCC), is similar to, but not exactly equivalent to, our trained decoder since our reward scheme, in addition to penalizing steps, also gives reward for annihilating syndrome defects. The latter will give a slight priority to using $Y$ operators (which can annihilate two pairs of defects) at an early stage of the decoding sequence. Nevertheless, we expect that this algorithm serves as a good benchmark for how well our \ac{DRL} implementation of the algorithm works. In particular, we would like to see that our decoder outperforms the MWPM decoder also for low error rates. 

\begin{figure*}[ht!]
\centering
\includegraphics[width=\linewidth]{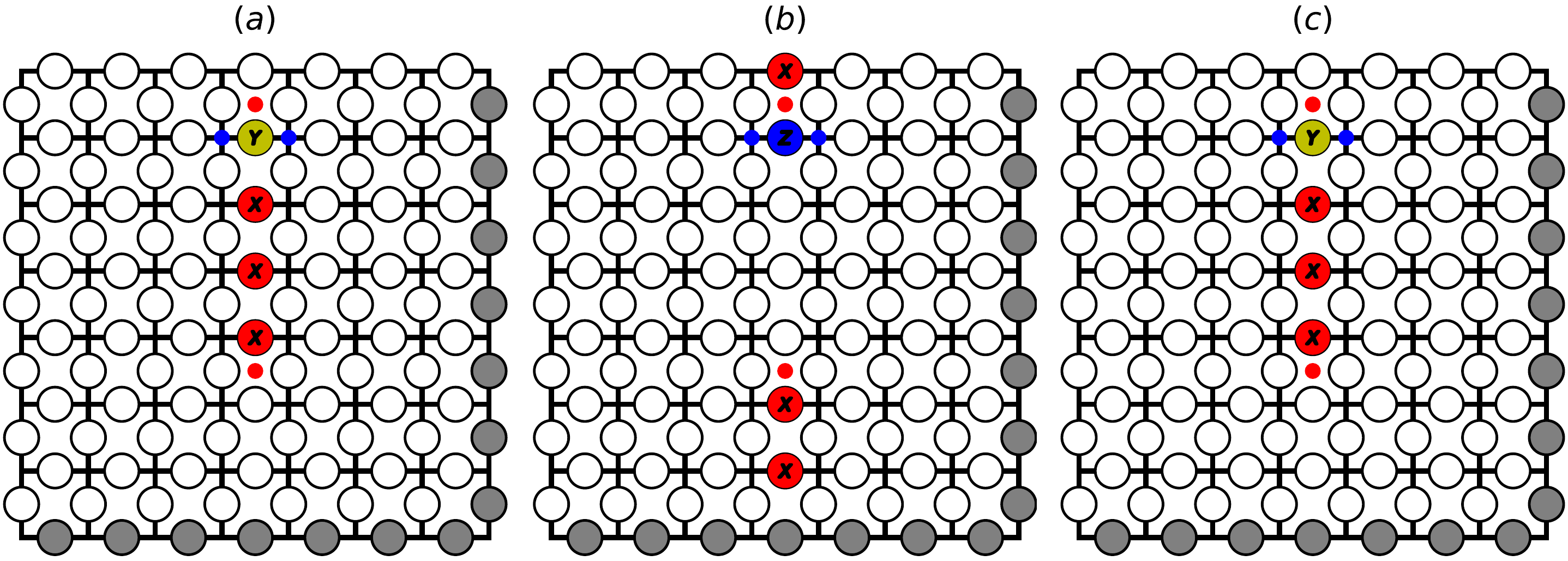}
\caption{(a) The initial syndrome corresponding to one $Y$ error and three $X$ errors. (b) \ac{MWPM} will always introduce a non-trivial loop and therefore fail. The ``minimum correction chain'' decoder has a $50\%$ probability each for failure and success [correction chains (b) or (c), respectively].
\label{fig:yxx}}
\end{figure*}

The shortest error strings that can give an error with either of the algorithms are $\lceil \frac{d}{2} \rceil$ long, aligned along one row or column~\cite{fowler2013optimal, andreasson2018quantum}. This means that the fail rate for both types of decoders will scale as $P_L\sim (p/3)^{\lceil \frac{d}{2} \rceil}$ for small $p$, but with different prefactors. We will only consider odd $d$; the scaling is true for even $d$, but prefactors are different. Figure~\ref{fig:yxx} gives a demonstrative example of an error string, for $d=7$, where the outcome differs between the two algorithms. Here MWPM will fail, solving the vertex defects with one $Z$ and the plaquette defects with two $X$ to generate a logical bit-flip consisting of a vertical $X$ loop. In contrast, the MCC algorithm will only fail 50\% of the time (we assume draws are settled by a coin flip), either using the MWPM-prescribed sequence or using the actual error string ($YXXX$) as the correction string. Interestingly, our specific decoder implementation should succeed 100\% of the time for this particular error string, since it will prefer to use the $Y$, but it is not clear that this advantage is general.

\begin{figure*}[ht!]
\centering
\includegraphics[width=\linewidth]{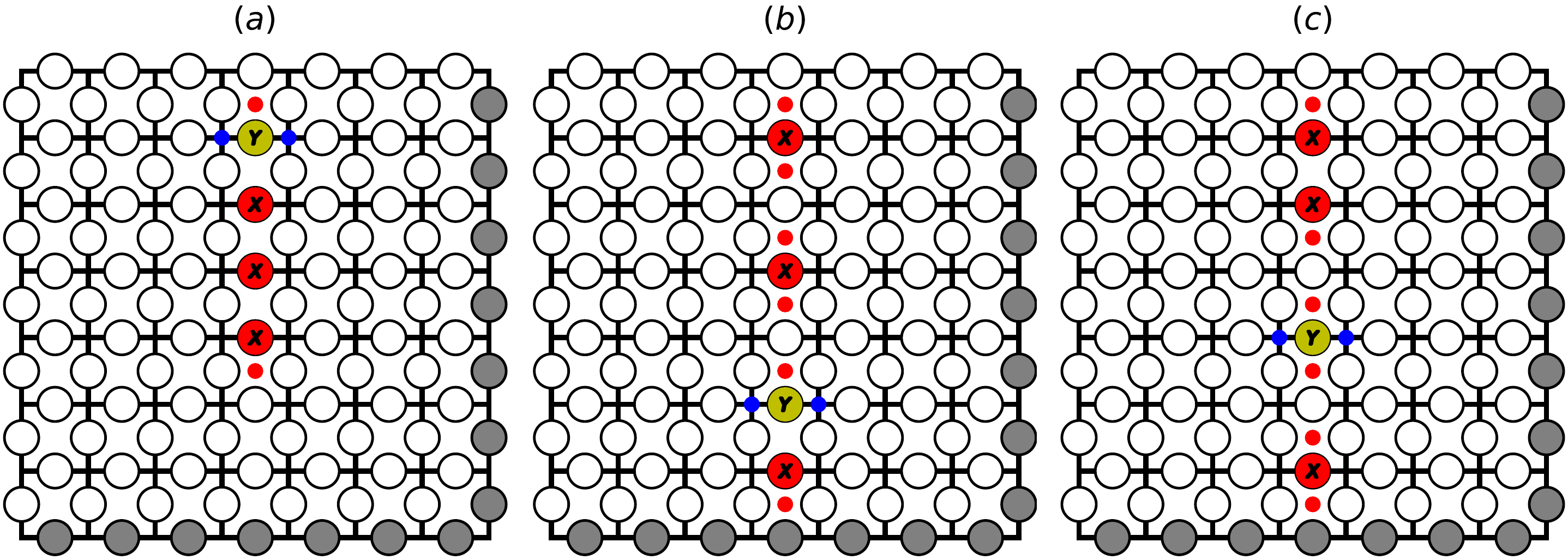}
\caption{(a-c) For each of these syndromes, the shortest correction chains are of the same length (four steps in all cases). This is also true for other constellations of errors. The length of the error correction chain does not depend on the relative position of the syndrome defects in a row or a column.
\label{fig:order_doesnt_matter}}
\end{figure*}

To derive the general expressions for the asymptotic fail rates, we go through several examples of error chains. First, one has to keep in mind that we are interested in the minimum amount of steps to annihilate all excitations. The order in which the errors are placed in the chain does not matter (see Fig.~\ref{fig:order_doesnt_matter}). Also, the errors do not have to be connected; it is a sufficient criterion that they all are in one column or row. 

\begin{figure*}[ht!]
\centering
\includegraphics[width=\linewidth]{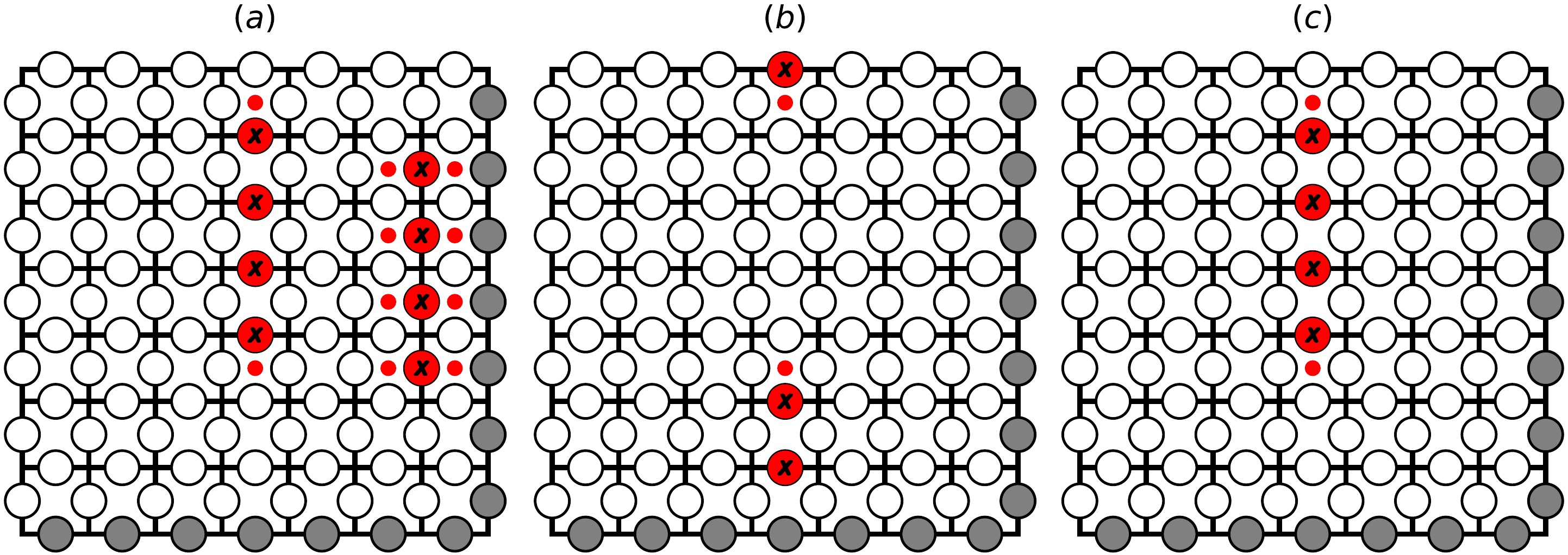}
\caption{(a) The initial syndrome with four $X$ errors. (b) The minimum amount of steps, three, to merge the excitations is by introducing a non-trivial loop around the torus. (c) Revoking the errors introduced would take four steps. Any decoder will fail on such error chains with 100$\%$ certainty. Note that $X$ chains of errors on the columns with vertical bonds, or rows with horizontal bonds, will not give QEC failure (panel a). 
\label{fig:only_x}}
\end{figure*}

\begin{figure*}[ht!]
\centering
\includegraphics[width=\linewidth]{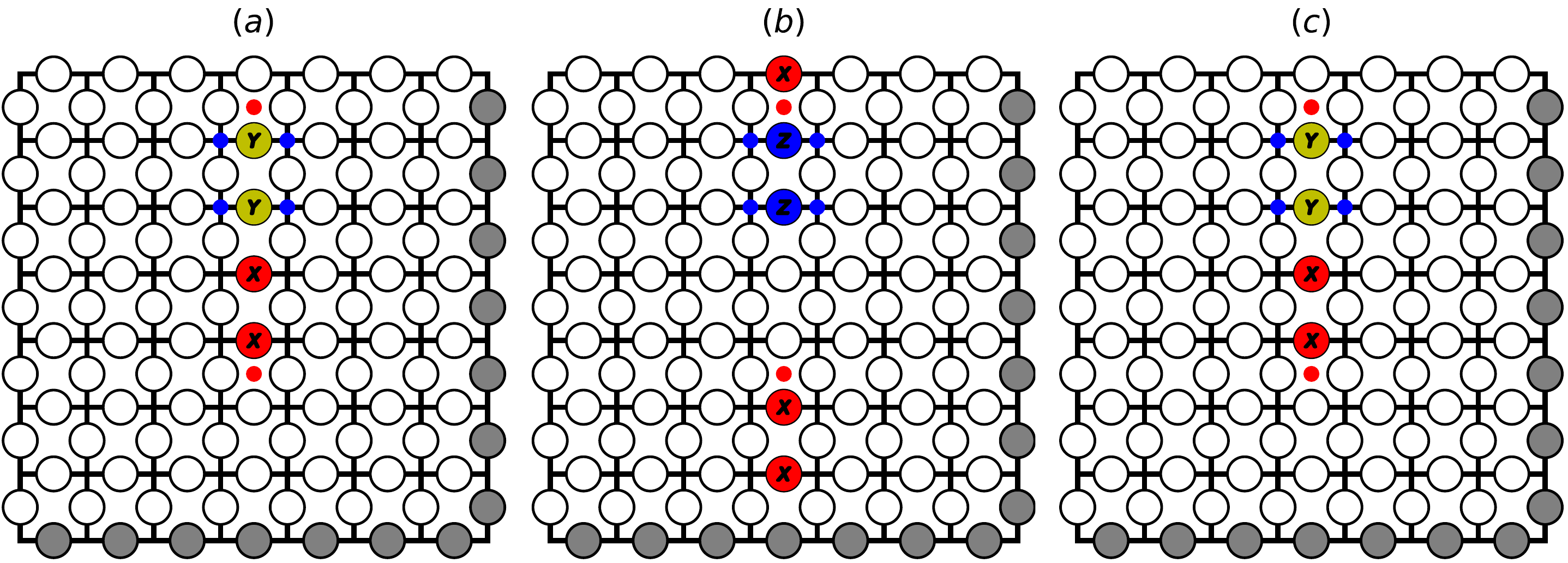}
\caption{(a) The initial syndrome with two $Y$ operators in the error chain. (b) Five steps are needed if one uses $Z$ operators. (c) There is only one shortest correction chain with four steps. We can also conclude that with at least two or more $Y$ errors in the chain, the MCC algorithm (and DRL decoder) always succeeds with the error correction. In contrast, MWPM will fail, using the middle chain (panel b).
\label{fig:yyxx}}
\end{figure*}

\begin{figure*}[ht!]
\centering
\includegraphics[width=\linewidth]{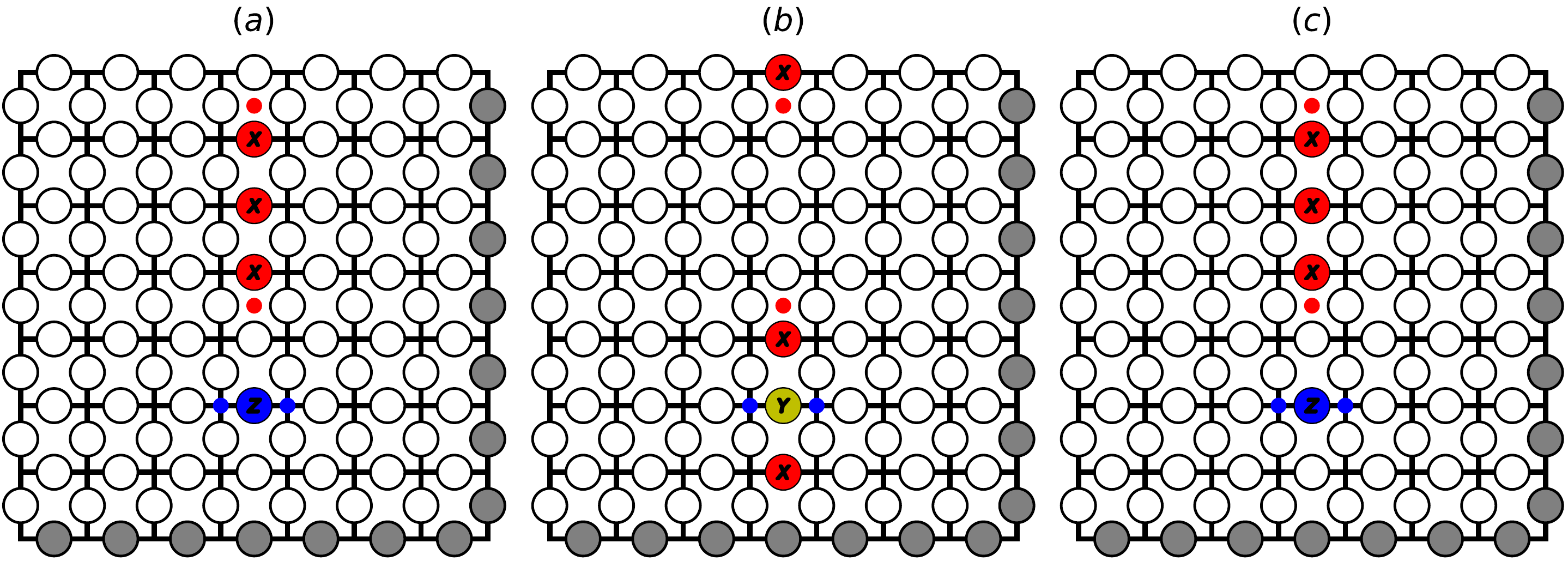}
\caption{(a) The initial syndrome with one $Z$ and three $X$ errors. There are two possible minimal error correction chains, one leading to (b) a failed and one leading to (c) a successful error correction. We assign 50$\%$ chance to each outcome. Interestingly, the \ac{MWPM} algorithm will always succeed on these kind of syndromes as $Y$ would count as two operators.
\label{fig:zxx}}
\end{figure*}

\begin{figure*}[ht!]
\centering
\includegraphics[width=\linewidth]{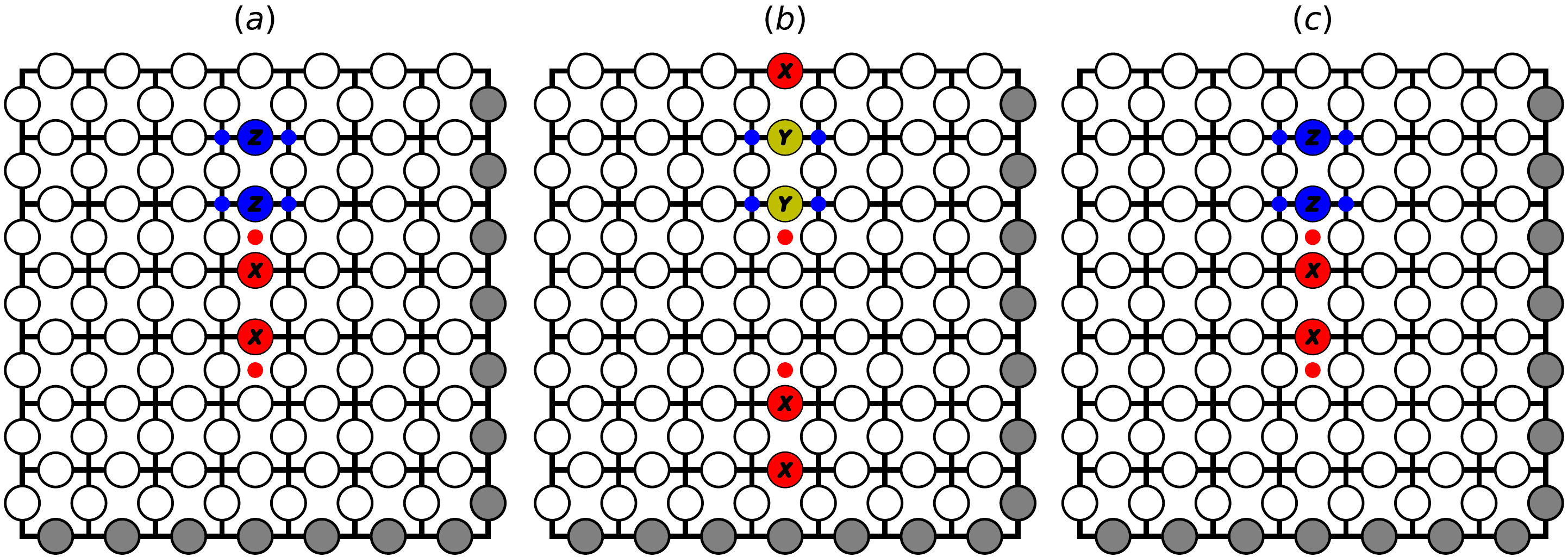}
\caption{The shortest error-correction chain for the initial syndrome is (a) four steps by simply (c) reversing the changes. (b) Using $Y$ operators would take five steps and will therefore not be chosen by the decoder. The agent always succeeds on these syndromes.
\label{fig:zzxx}}
\end{figure*}

Now we can investigate the different combinations that can make the decoder fail. Length $\lceil \frac{d}{2} \rceil$ error chains containing either only $X$ or $Z$ errors will always generate a non-trivial loop (see Fig.~\ref{fig:only_x}). Moreover, combinations of $X$ and $Y$ errors can lead to a failure. Figures \ref{fig:yxx} and \ref{fig:yyxx} show that we have to consider syndromes with exactly one $Y$ error and the rest uniformly $X$ or $Z$ errors. For two or more $Y$ errors, the decoder will always succeed with the error correction. Finally, we have to find out how $X$ and $Z$ errors in combination behave. Figures \ref{fig:zxx} and \ref{fig:zzxx} show that for exactly one $Z$ error and the rest being $X$ errors, the decoder succeeds with a $50\%$ chance. Here again, the reward scheme of the actual \ac{DRL} decoder would disfavor using a $Y$ if the $Z$ is isolated, giving a slight discrepancy between this and the MCC algorithm.


We can convince ourselves that the cases presented here generalize to larger odd $d$, allowing for the derivation of an analytic expression for the logical fail rate. For the MCC algorithm, which we identify as close to the performance of our \ac{DRL} decoder, the fail rate is given by
\begin{eqnarray}
    P_{L_{MCC}} &=& P(\{XX \ldots X\}) + P(\{ZZ \ldots Z\}) \nonumber\\
    &&+ P(\{YX \ldots X\}) + P(\{YZ \ldots Z\}) \nonumber\\
    &&+ P(\{ZX \ldots X\}) + P(\{XZ \ldots Z\}),
    \label{equ:p_l_RL}
\end{eqnarray}
where $\{ \ldots \}$ indicates any configuration of errors in one row or column.  

To lowest order in $p$ [i.e., ignoring factors that are powers of $(1-p)$], the probability of $\lceil \frac{d}{2}\rceil$ errors of the same type is given by 
\begin{equation}
     P(\{XX \ldots X\}) = P(\{ZZ \ldots Z\}) = 2d \cdot \binom{d}{\lceil \frac{d}{2}  \rceil} \cdot \mleft( \frac{p}{3} \mright)^{\lceil \frac{d}{2}\rceil},
     \label{equ:p_xxx}
\end{equation}
where the $2d$ corresponds to the number of rows and columns (with the appropriate orientation of bonds; see Fig.~\ref{fig:only_x}). The probability of failure from the mixed-type chains is given by 
\begin{eqnarray}
    &&P(\{YX \ldots X\}) = P(\{YZ \ldots Z\}) \nonumber\\
    &=& P(\{ZX \ldots X\}) = P(\{XZ \ldots Z\}) \nonumber\\
    &=& \frac{1}{2} 2d \cdot \binom{d}{1} \cdot \mleft( \frac{p}{3} \mright) \cdot \binom{d-1}{\lceil\frac{d}{2}\rceil-1} \cdot \mleft( \frac{p}{3} \mright)^{\lceil \frac{d}{2} \rceil - 1}\nonumber\\
    &=& d\,\lceil \frac{d}{2} \rceil \cdot \binom{d}{\lceil \frac{d}{2} \rceil} \cdot \mleft( \frac{p}{3} \mright)^{\lceil \frac{d}{2} \rceil},
    \label{equ:p_yxx}
\end{eqnarray}
where the $\frac{1}{2}$ comes from 50\% failure for this type of configuration. 
Inserting Eqs.~(\ref{equ:p_xxx}) and (\ref{equ:p_yxx}) in Eq.~(\ref{equ:p_l_RL}) and simplifying, we obtain the following probability of failure in the case of very low $p$:
\begin{equation}
    P_{L_{MCC}} =
    4 d (1+\lceil\frac{d}{2}\rceil)\,\binom{d}{\lceil\frac{d}{2}\rceil}\cdot \Big( \frac{p}{3} \Big) ^{\lceil\frac{d}{2}\rceil} 
\end{equation}

To derive the corresponding asymptotic fail rate for the MWPM algorithm, we use the fact that it only uses $X$ and $Z$ for correction. This decoder (similarly to any reasonable decoder) will always fail for chains of length $\lceil \frac{d}{2}  \rceil$ in a row or column containing all $X$ or all $Z$. It will also fail if one or more of the $X$ or $Z$ in such a chain are replaced by $Y$. This is clear from, e.g., correcting a $Y$ with a $Z$ in a chain $\{YXX \ldots \}$, which will reduce the chain to a pure $\{XXX \ldots \}$ of the type that always fails.
\begin{eqnarray}
    P_{L_{MWPM}} &=& P(\{XX \ldots X\}) + P(\{ZZ \ldots Z\}) \nonumber\\
    &&+ P(\{YX \ldots X\}) + P(\{YZ \ldots Z\}) \nonumber\\
    &&+ \ldots + P(\{YY \ldots Y\}),
\end{eqnarray}
where the ellipsis indicates chains with increasing numbers of $Y$. 
The general expression for $N_y \in \{0,1,\cdots,\lceil\frac{d}{2}\rceil\}$ $Y$ errors in a chain with $\lceil \frac{d}{2} \rceil - N_y$ $X$ ($Z$) errors reads
\begin{eqnarray}
    &&P(\{YY...XX\}) = P(\{YY...ZZ\}) \nonumber\\
    &=& 2 \mleft( 1 + \delta_{N_y, \lceil \frac{d}{2} \rceil} \mright) d \cdot \binom{d}{N_y} \cdot \binom{d - N_y}{\lceil \frac{d}{2} \rceil - N_y} \cdot \mleft( \frac{p}{3} \mright)^{\lceil \frac{d}{2} \rceil}, \qquad
    \label{equ:p_yyx}
\end{eqnarray}
where, compared to Eq.~(\ref{equ:p_yxx}), there is no $\frac{1}{2}$, as these chains always fail using MWPM, and where the chain consisting purely of $Y$ is multiplied by a factor of 2 because it will fail on both types ($X$ or $Z$) of rows and columns. 
Thus, the complete expression for the MWPM asymptotic fail rate reads (after summation over $N_y$)
\begin{equation}
    P_{L_{MWPM}} =
    4 d \cdot 2^{\lceil\frac{d}{2}\rceil} \binom{d}{\lceil\frac{d}{2}\rceil} \left(\frac{p}{3}\right)^{\lceil \frac{d}{2}\rceil}.
    \label{eqn:MWPM_fail}
\end{equation}

As expected, we find a higher fail rate for the decoder that uses MWPM compared to the decoder using the minimum number of correction steps, with $P_L / P_{L_{MWPM}} = (1 + \lceil \frac{d}{2} \rceil) / 2^{\lceil \frac{d}{2} \rceil} < 1$ for $d \geq 3$. 

We also note that the asymptotic fail rate for pure bit-flip (or phase-flip) noise with error rate $p$ is given by Eq.~(\ref{equ:p_xxx}) with $p/3 \rightarrow p$, $P_{L,X}(p) = 2d \cdot \binom{d}{\lceil \frac{d}{2} \rceil} \cdot p^{\lceil \frac{d}{2} \rceil}$. Thus, under the assumption of uncorrelated $X$ and $Z$ errors with probability $2p/3$ (corresponding to the rates for depolarizing noise) we find exactly that the total fail rate in Eq.~(\ref{eqn:MWPM_fail}) is given by adding up two independent error channels: $P_{L_{MWPM}} = 2 P_{L,X} (2p/3)$. 

Another useful representation is to calculate the ratio of error chains with $\left\lceil \frac{d}{2} \right\rceil$ errors that lead to a failure compared to the total number of chains with $\left\lceil \frac{d}{2} \right\rceil$ errors: 
\begin{align}
    f_{RL} = \frac{4 d \cdot (1+\lceil\frac{d}{2}\rceil)\,\binom{d}{\lceil\frac{d}{2}\rceil}}{\binom{2d^2}{\lceil\frac{d}{2}\rceil}\cdot \lceil \frac{d}{2} \rceil^3}.
\end{align}
Accordingly, for the MWPM: 
\begin{align}
    f_{MWPM} = \frac{4 d \cdot 2^{\lceil\frac{d}{2}\rceil} \binom{d}{\lceil\frac{d}{2}\rceil}}{\binom{2d^2}{\lceil\frac{d}{2}\rceil}\cdot \lceil \frac{d}{2} \rceil^3}.
\end{align}

\section{Model definition, hyperparameters, and running time}
\label{appendix:model_definition_hyperparameters}

In this section, we list relevant parameters for our neural networks. Table \ref{tab:Hyperparameter} shows the different hyperparameters used in training along with short descriptions of each. The structure of the deep neural network used for most of the training can be seen in Tables~\ref{tab:NN1} and  \ref{tab:NN2}. The network consists of mostly convolutional 2-dimensional layers of decreasing size. All layers except the first used zero-padding. The first layer used padding with periodic boundary conditions. For grid size $d=9$, we used the built-in ResNet34 definition provided in the PyTorch framework. It has 21,277,955 tunable parameters.

\begin{table*}[ht!]
	\centering
	\renewcommand{\arraystretch}{1.3}
	\caption{List of hyperparameters and their values.}
	\label{tab:Hyperparameter}
	\begin{tabular}{p{5cm} p{2.5cm} p{8cm}}
		\hline
		Hyperparameter & Value & Description \\
		\hline
		mini-batch size & 32 &  Number of training samples used for stochastic gradient descent update.\\
		training steps & 10000 & Total amount of training steps per epoch.\\
		replay memory size, $N$ & 10000 & Total amount of stored memory samples.\\
		priority exponent, $\alpha$ & $0.6$ & Prioritized experience replay parameter. \\
		importance weight, $\beta$ & $0.4$ & Prioritized experience replay parameter. \\
		target network update frequency, $C$ & 1000 & The frequency with which the target network is updated with the policy network. \\
		discount factor, $\gamma$ & 0.95 & Discount factor $\gamma$ used in the Q-learning update.\\
		learning rate & 0.00025 & The learning rate used by Adam.\\
		initial exploration & 1 & Initial value of $\epsilon$ in $\epsilon$-greedy exploration. \\
		final exploration & 0.1 & Final value of $\epsilon$ in $\epsilon$-greedy exploration.\\
		replay start size & 1000 & A random policy generates training samples to populate the replay memory before the learning starts.\\
		optimizer & Adam & Adam is an optimization algorithm used to update network weights.\\
		max steps per episode & 75 & Number of steps before every episode is terminated.\\
		\hline
	\end{tabular}
\end{table*}

\begin{table}[ht!]
\centering
\caption{Network architecture for $d=5$. Every convolutional layer has a kernel size of 3 and stride 1. Periodic padding is applied to the first convolutional layer. The other convolutional layers work with zero-padding.}
\label{tab:NN1}
\begin{tabular}{p{0.5cm} p{1.5cm} p{1.5cm} p{2cm}}
		\hline
		$\#$ & Type & Size & $\#$parameters \\
		\hline
		1 & Conv2d & 128 & 2,432\\
		2 & Conv2d & 128 & 147,584\\
		3 & Conv2d & 120 & 138,360\\
		4 & Conv2d & 111 & 119,991\\
		5 & Conv2d & 104 & 104,000\\
		6 & Conv2d & 103 & 96,511\\
		7 & Conv2d & 90 & 83,520\\
		8 & Conv2d & 80 & 64,880\\
		9 & Conv2d & 73 & 52,633\\
		10 & Conv2d & 71 & 46,718\\
		11 & Conv2d & 64 & 40,960\\
		12 & Linear & 3 & 1,731\\
		\hline
		 &  &  & 899,320\\
		\hline
	\end{tabular}
\end{table}

\begin{table}[ht!]
\centering
\caption{Network architecture for $d=7$. Every convolutional layer has a kernel size of 3 and stride 1. Periodic padding is applied to the first convolutional layer. The other convolutional layers work with zero-padding.}
\label{tab:NN2}
\begin{tabular}{p{0.5cm} p{1.5cm} p{1.5cm} p{2cm}}
		\hline
		$\#$ & Type & Size & $\#$parameters \\
		\hline
		1 & Conv2d & 256 & 4,864\\
		2 & Conv2d & 256 & 590,080\\
		3 & Conv2d & 251 & 578,555\\
		4 & Conv2d & 250 & 565,000\\
		5 & Conv2d & 240 & 540,240\\
		6 & Conv2d & 240 & 518,640\\
		7 & Conv2d & 235 & 507,835\\
		8 & Conv2d & 233 & 493,028\\
		9 & Conv2d & 233 & 488,834\\
		10 & Conv2d & 229 & 480,442\\
		11 & Conv2d & 225 & 463,950\\
		12 & Conv2d & 223 & 451,798\\
		13 & Conv2d & 220 & 441,760\\
		14 & Conv2d & 220 & 435,820\\
		15 & Conv2d & 220 & 435,820\\
		16 & Conv2d & 215 & 425,915\\
		17 & Conv2d & 214 & 414,304\\
		18 & Conv2d & 205 & 395,035\\
		19 & Conv2d & 204 & 376,584\\
		20 & Conv2d & 200 & 367,400\\
		21 & Linear & 3 & 15,003\\
		\hline
		 &  &  & 8,990,907\\
		\hline
	\end{tabular}
\end{table}

The hardware used for the training was one GPU unit (NVIDIA Tesla V100 SMX2 GPU). The training time depends on the grid size. The bigger the grid, the more training is necessary. With the implementation found on github, $d=5$ converged after 5 hours of training. The network for $d=7$ needs approximately 4 days (96 hours) for convergence. 

\section{Selected episodes}
\label{appendix:selected_episodes}

\begin{figure*}[ht!]
\centering
\includegraphics[width=0.74\textwidth]{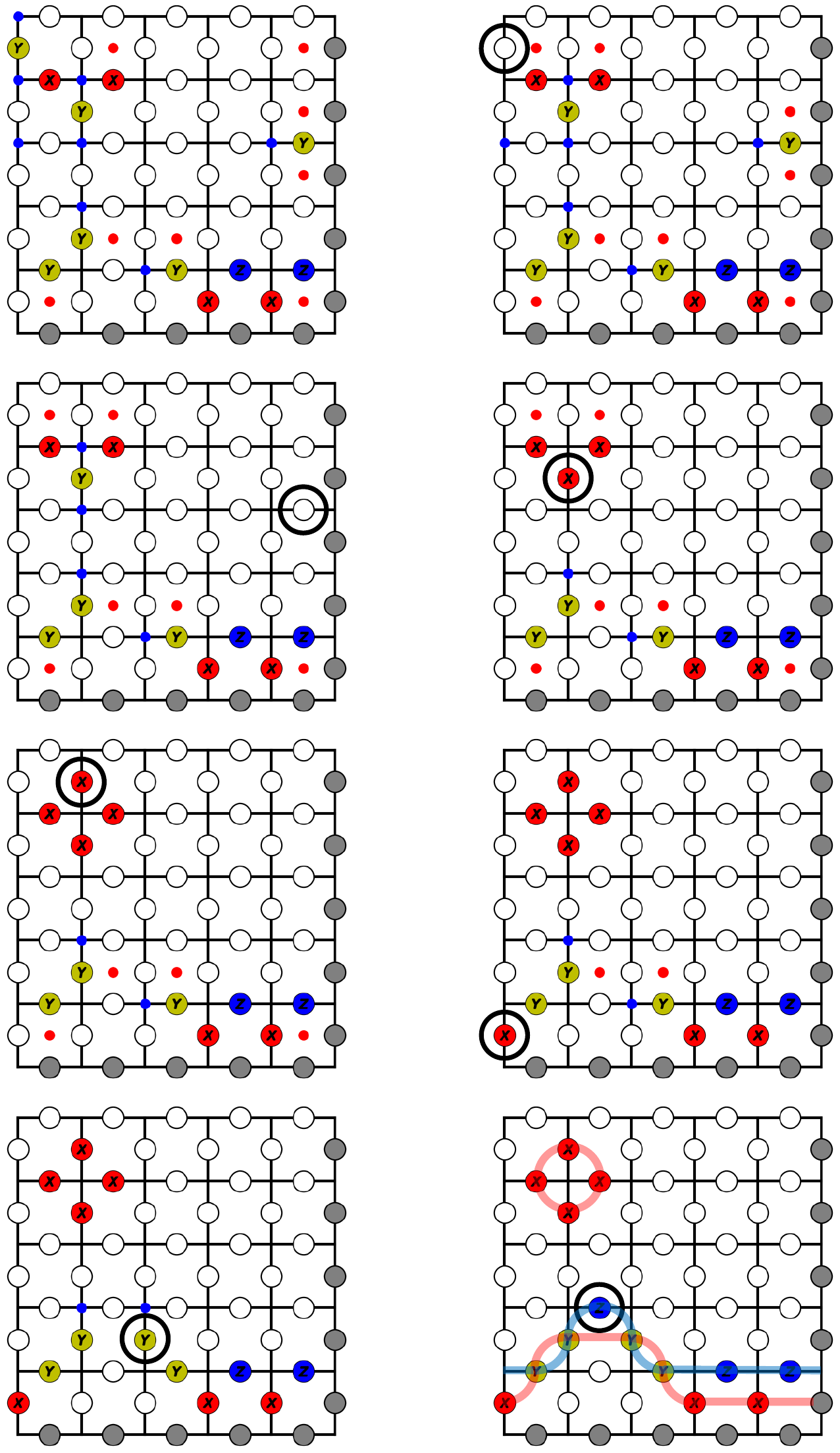}
\caption{A selected correction sequence from the fully trained decoder. The sequence goes from left to right and top to bottom. The circles indicate on which qubit an action was performed. In this case, the error correction fails, with the last state corresponding to a logical $Y$-operator, i.e.\ both bit- and and phase-flip.
\label{fig:sequence_correction_fail}}
\end{figure*}

\begin{figure*}[ht!]
\centering
\includegraphics[width=0.74\textwidth]{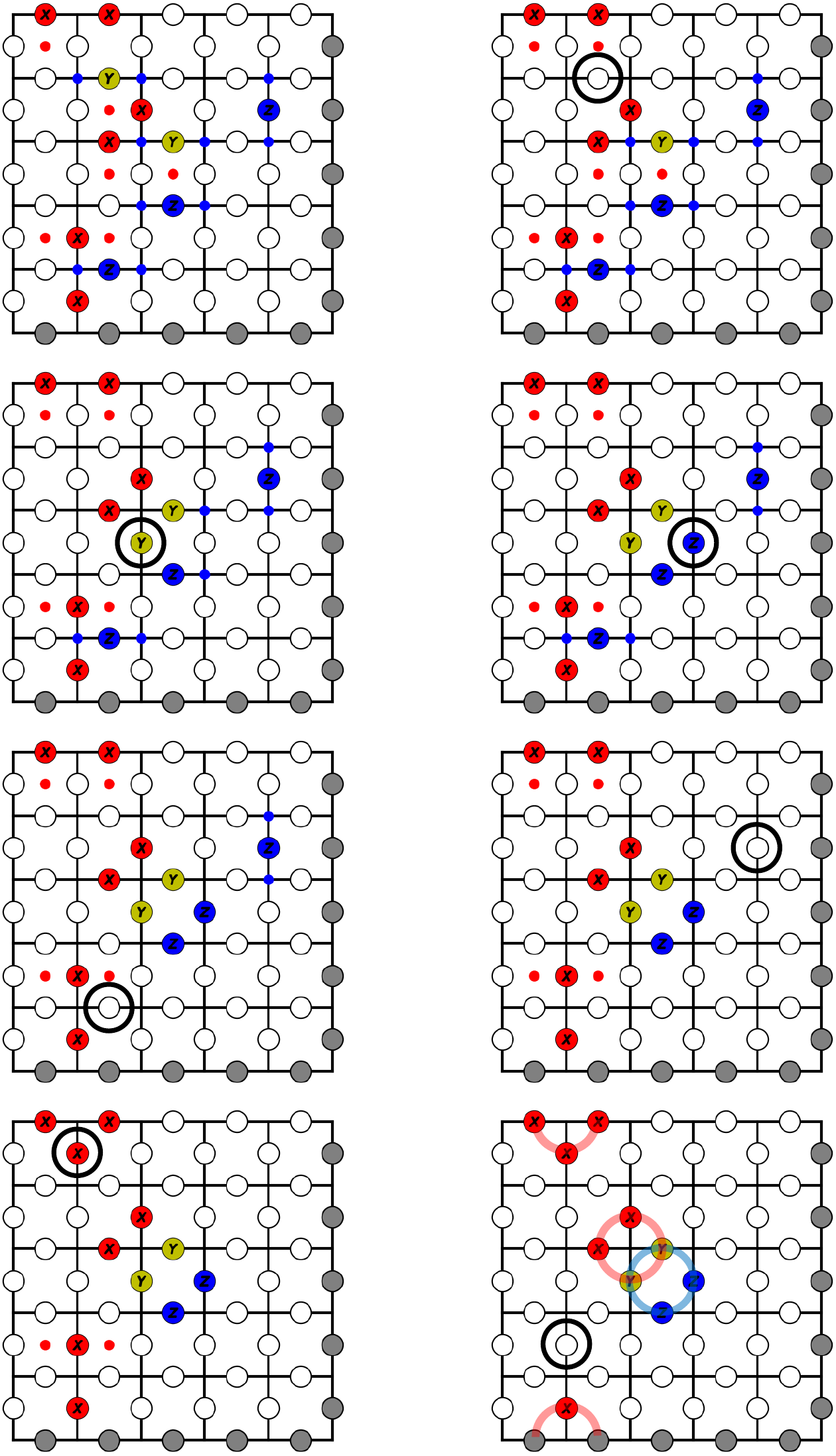}
\caption{A selected correction chain from the fully trained decoder. The sequence goes from left to right and top to bottom. The circles indicate on which qubit an action was performed. Here the error correction is successful, with only trivial loops remaining.
\label{fig:sequence_correction_success}}
\end{figure*}

In this section, we present two selected episodes of error correction using the fully trained decoder for $d=5$. Figure~\ref{fig:sequence_correction_fail} shows an example where the error correction fails and Fig.~\ref{fig:sequence_correction_success} shows an example of succesful error correction.

\clearpage

\end{document}